\documentclass[aps,pra,twocolumn,groupedaddress,amsmath,amssymb,floatfix,nofootinbib]{revtex4}

\usepackage{graphics}      
\usepackage{longtable}     
\usepackage{url}           
\usepackage{bm}            
\usepackage{bbm}
\usepackage{dcolumn}
\usepackage{subfigure}
\usepackage{pdfsync}
\usepackage{appendix}

 \usepackage[pdftex]{graphicx}

  \usepackage[colorlinks=true,bookmarks=true,%
     pdfborder={0 0 0},linkcolor=blue,urlcolor=red,%
     breaklinks=true,bookmarksnumbered=true]{hyperref}

\long\def\symbolfootnote[#1]#2{\begingroup%
\def\thefootnote{\fnsymbol{footnote}}\footnote[#1]{#2}\endgroup}

\def \x{\bm{x}}
\def \y{\bm{y}}
\def \z{\bm{z}}
\def \r{\bm{r}}
\def \S{\bm{S}}
\def \T{\bm{T}}
\def \k{\bm{k}}
\def \p{\bm{p}}
\def \q{\bm{q}}
\def \mC{\mathcal{C}}
\def \n{\hat{\bm{n}}}
\def \beq{\begin{equation}}
\def \eeq{\end{equation}}
\def \beqarr{\begin{eqnarray}}
\def \eeqarr{\end{eqnarray}}
\def \bspt{\begin{split}}
\def \espt{\end{split}}
\def \bef{\begin{figure}}
\def \enf{\end{figure}}
\def \tr{\text{tr }}
\def \bpm{\begin{pmatrix}}
\def \epm{\end{pmatrix}}
\def \a{\hat{a}}
\def \b{\hat{b}}

\def \pt{\partial}

\newcommand{\abs}[1]{\lvert#1\rvert}

\newcommand{\meanvalue}[3]{\mbox{$\langle #1 | #2 | #3 \rangle$}}

\newcommand{\ev}[1]{\mbox{$\langle #1 \rangle$}}

\newcommand{\ket}[1]{\mbox{$| #1 \rangle$}}

\begin{document}

\title{Entanglement Entropy of Fermi Liquids via Multi-dimensional Bosonization}

\author{Wenxin Ding$^{1}$}
\author{Alexander Seidel$^{2}$}
\affiliation{}
\author{Kun Yang$^{1}$}
\affiliation{(1) National High Magnetic Field Laboratory and
  Department of Physics, Florida State University, Tallahassee, FL
  32306, USA \\ (2) Department of Physics and Center for Materials
  Innovation, Washington University, St. Louis, MO 63136, USA}

\date{\today}

\begin{abstract}
The logarithmic violations of the area law, i.e. an ``area law" with logarithmic correction of the form $S \sim L^{d-1} \log L$, for entanglement entropy are found in both 1D gapless fermionic systems with Fermi points and for high dimensional free fermions. The purpose of this work is to show that both violations are of the same origin, and in the presence of Fermi liquid interactions such behavior persists for 2D fermion systems. In this paper we first consider the entanglement entropy of a toy model, namely a set of decoupled 1D chains of free spinless fermions, to relate both violations in an intuitive way. We then use multi-dimensional bosonization to re-derive the formula by Gioev and Klich [Phys. Rev. Lett. 96, 100503 (2006)] for free fermions through a low-energy effective Hamiltonian, and explicitly show the logarithmic corrections to the area law in both cases share the same origin: the discontinuity at the Fermi surface (points). In the presence of Fermi liquid (forward scattering) interactions, the bosonized theory remains quadratic in terms of the original local degrees of freedom, and after regularizing the theory with a mass term we are able to calculate the entanglement entropy perturbatively up to second order in powers of the coupling parameter for a special geometry via the replica trick. We show that these interactions do not change the leading scaling behavior for the entanglement entropy of a Fermi liquid. At higher orders, we argue that this should remain true through a scaling analysis.
\end{abstract}

\maketitle

\section{Introduction}
The study of entanglement, which is one of the most fundamental aspects of quantum mechanics, has lead to and is still leading to much important progress and applications in different fields of modern physics such as quantum information\cite{nielsen2000}, condensed matter physics\cite{Amico2009,Latorre2009,Eisert2010}, etc..  To name a few, it has lead to better undertanding of density matrix renormalization group (DMRG)\cite{White1992, Verstraete2004, Perez-Garcia2006}; it has also been proposed to be a tool for the  characterization of certain topological phases\cite{levin:110405,kitaev:110404,hli:2008}.

Among various ways of quantifying entanglement, in condensed matter or many-body physics efforts have mainly focused on the bipartite block entanglement entropy (von Neumann entropy) and  its generalizations (R\'{e}nyi or Tsallis entropy). It has become increasingly useful in characterizing phases\cite{Chen2011} and phase transitions\cite{Osborne2002,PhysRevLett.90.227902}. The area law\cite{area-law} is one of the most important results on entanglement entropy: it states that the entanglement entropy is proportional to the area of the surface separating two subsystems. However, thus far there are two important classes of systems that violate the area law: in gapless one dimensional (1D) systems, a logarithmic divergence\cite{wilczek94, PhysRevLett.90.227902} is found where according to the area law the entanglement entropy should saturate as the size of the subsystem grows; in higher dimensions, for free fermions the area law is found to be corrected by a similar logarithmic factor $\log L$\cite{gioev:100503,wolf:010404,Cramer07,WFLi06,Barthel06,Levine08,YZhang2011,Helling2011}, where $L$ is the linear dimension of the subsystem.

In this work, we first show that
the scaling behavior of the entanglement entropy for systems with a Fermi surface is the same as that of 1D systems with Fermi points\cite{footnote,Ding09,Swingle2010,Swingle2}.
We then seek for a generalization of the latter to interacting fermions in the Fermi liquid phase. We first develop an intuitive understanding via a toy model, showing that in this model the entanglement entropy has the same form as that given in by Gioev and Klich (GK) in Ref.[\onlinecite{gioev:100503}]. We then develop a more general and formal treatment using the method of high-dimensional bosonization\cite{houghton1993, haldane1993, CastroNeto1994, CastroNeto1994a, CastroNeto1995}. This approach will not only lead to a reproduction of the result for free fermions obtained by GK based on Widom's conjecture\cite{widom1982,sobolev}, but will also lend itself to the inclusion and subsequent treatment of Fermi liquid type (forward scattering) interactions.

This paper is organized as follows. In Sec. \ref{sec:toy_model}, we describe the toy model for which the entanglement entropy can be written in the same form as the GK result. Then in Sec. \ref{sec:bosonization}, we briefly introduce the tool box of multi-dimensional bosonization, and apply it to free fermions to reproduce the GK formula. The main results of this work are presented in Sec. \ref{sec:ee} in which we calculate the entanglement entropy of a Fermi liquid for a special geometry using a combination of multi-dimensional bosonization and the replica trick. We subsequently summarize and discuss our results. Some technical details are discussed in two appendices.

\section{The Intuitive picture - a toy model}\label{sec:toy_model}

\begin{figure}
\centering
  \subfigure[]{
\label{fig:realspace}
\includegraphics[width=7cm]{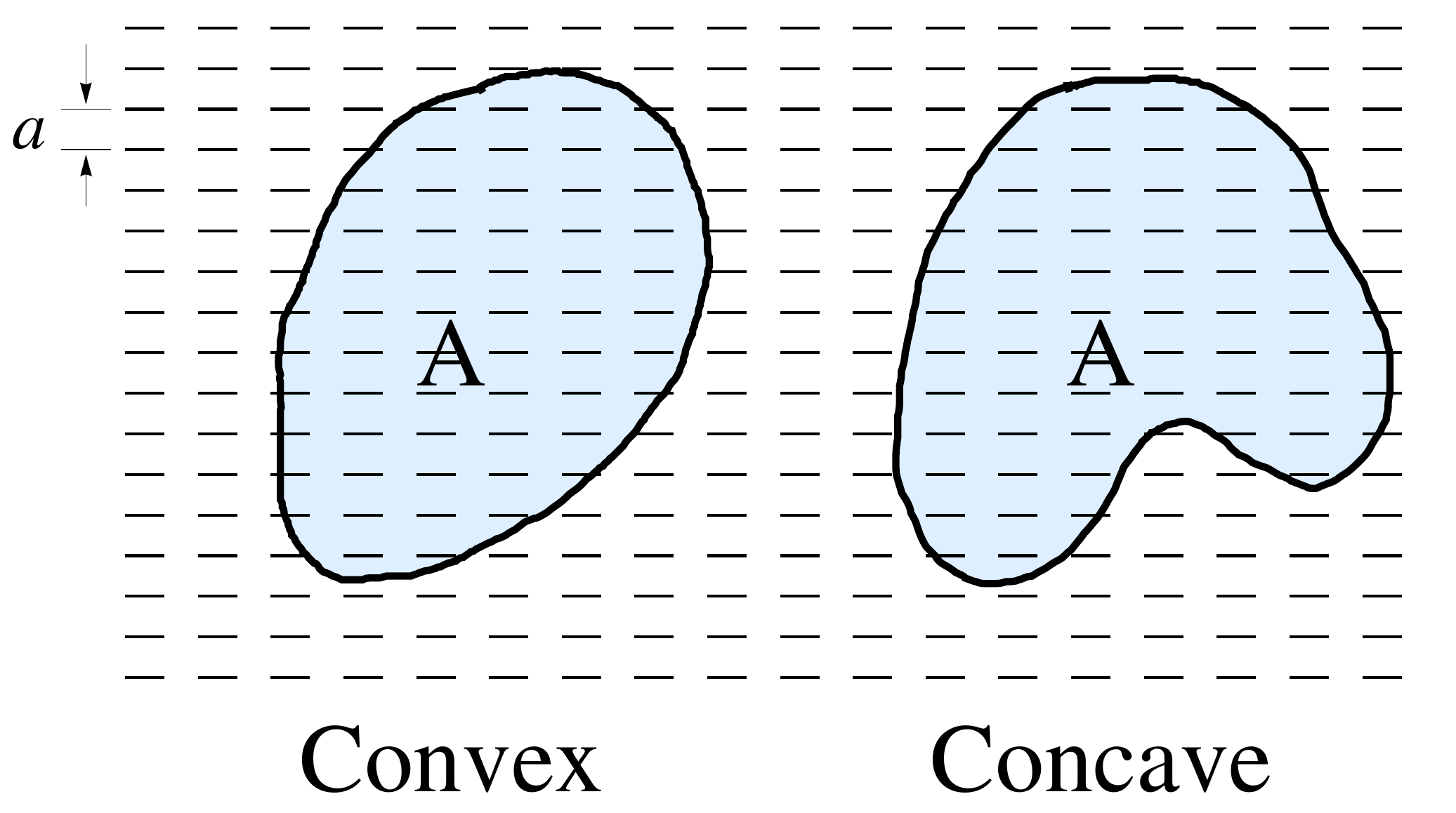}}
\centering
  \subfigure[]{
\label{fig:fermisurface}
\includegraphics[width=7cm]{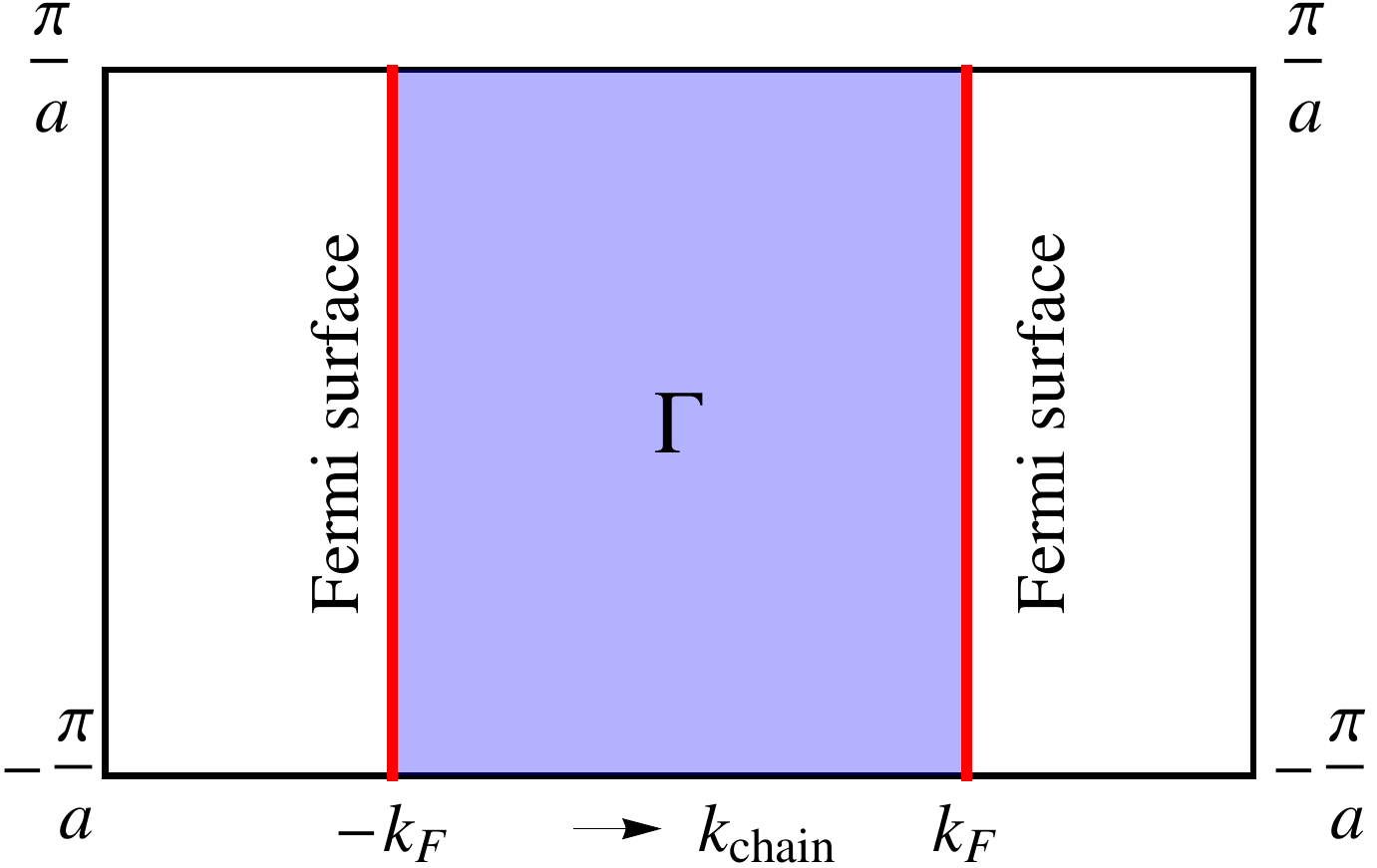}}
\caption{(Color online) The toy model both in real space and in momentum space. (a) A set of parallel decoupled 1$d$ chains of spinless free fermions (dash lines); the subsystem division is represented by the solid lines, both convex and concave geometries. (b) Fermi surfaces of the toy model.}
  \label{fig:toy_model} 
\end{figure}

Consider a set of decoupled parallel 1D chains of non-interacting spinless fermions with spacing $a$ as shown in Fig. \ref{fig:realspace}. Here we only consider $d=2$ for simplicity, but this toy model is viable in general $d$ dimensions. The asymptotic behavior of entanglement entropy in large $L$ limit of a convex subsystem $A$ of  this model can be obtained by simply counting the number of chains that intersect $A$, and each segment contributes a $(1/3) \log L$ where $L$ is the linear dimension of the subsystem\cite{area-law,Korepin2004}. Due to the logarithm, different shapes only lead to differences at the area law level. Since each segment must have two intersections, we can count the intersections instead, which also automatically takes care of non-convex geometries. Although there is an additional correction for multiple intervals on a single chain\cite{Calabrese2009b, Alba2011, furukawa2009}, as long as only the $\log L$ behavior is concerned, that contribution is negligible. For $L$ large enough, we can write the number of these intersections as an integral over the surface of $A$ projected onto the direction perpendicular to the chains times one half of the chain density, $1/a$. To make contact with the GK result, we note this model also has Fermi surfaces as shown in Fig. \ref{fig:fermisurface} with a total ``area'' of $4\pi / a$. This enables us to replace the density of chains by an integral over the Fermi surfaces of the system $$\frac{1}{a} = \frac{1}{4\pi} \oint_{\pt \Gamma} dS_k,$$ where $\Gamma$ indicates the occupied area in momentum space so its boundary $\pt \Gamma$ is the Fermi surface(s). Therefore we can write the entanglement entropy as
\beq\label{eq:toy_model}
\begin{split}
S(\rho_A) &= \frac{1}{2} \times \frac{1}{3} \log L \times \frac{1}{a} \oint_{\partial A}
\abs{\n \cdot dS_x} \\
&= \frac{1}{12 (2\pi)^{2-1}} \log L \times
\oint_{\partial A} \oint_{\partial \Gamma} \abs{dS_x \cdot dS_k},
\end{split}
\eeq
where $\n$ is the direction along the chains which is also normal to the Fermi surface, and an overall factor of $\frac 12$ accounts for the double counting of chain segments. In Eq. (\ref{eq:toy_model}) we recover the GK formula in this special case but written in a slightly different way. In Ref. [\onlinecite{gioev:100503}] the entanglement entropy is given as:
\beq
S = \frac{1}{12}\frac{L^{d-1}\log L}{(2\pi)^{d-1}} \oint_{\partial A}
\oint_{\partial \Gamma} \abs{\n_{\x} \cdot \n_{\p}} dS_{\x} dS_{\p},
\eeq
where the real space surface integral is carried out over the subsystem whose volume is normalized to $1$. The surface area is factored out as $L^{d-1}$. However, in our formula the surface area $\sim L^{d-1}$ is implicitly included in the integral over the surface of the subsystem.

We note that the model discussed in Ref.\cite{Swingle2010} is equivalent with our toy model, but motivated from a different perspective. In Ref.\cite{Swingle2010}, models are constructed from the momentum space, either with Fermi surfaces as our toy model, or a square Fermi surface, and a boxlike and a spherical geometry are discussed. In contrast, our toy model is constructed from a real space perspective, and general single connected geometries are discussed.

Motivated by the toy model, in this work we extend this intuitive understanding of GK's result to generic free Fermi systems and generalize it to include Fermi liquid interactions in two dimensions (2D) via high dimensional bosonization. Using the method of multi-dimensional bosonization, the Fermi liquid theory can be written as a tensor product of  low-energy effective theories of quasi-1D systems similar to this toy model, along all directions. This provides us with a tool to treat the entanglement entropy of fermions in high dimensions, even in the presence of {\it interactions}.

At this point, we could also include forward scattering for each chain, and from 1D bosonization we know that for spinless fermions this only leads to renormalization of the Fermi velocity, thus does not change the logarithmic scaling of the entanglement entropy for this toy model. This hints that the same conclusion might hold for Fermi liquids, as we can include Fermi liquid interactions in a similar way via high dimensional bosonization. Although as we show later, this is indeed true at the leading order, the situation is more delicate than it seems to be. The Fermi liquid interactions couple a family of ``toy models'' aligned along different directions in the language of high dimensional bosonization, and lead to a correction to the entanglement entropy $\sim \mathcal{O}(1) \times \log L$.

\section{Multi-dimensional bosonization}\label{sec:bosonization}
The scheme of multi-dimensional bosonization was first introduced by Haldane\cite{haldane1993}, followed by others\cite{houghton1993, CastroNeto1994, CastroNeto1994a,  CastroNeto1995}. The basic idea is to start with a low energy effective Hamiltonian (obtained through a renormalization group (RG) approach) restricted to within a thin shell of thickness $\lambda$ around the Fermi surface, $k_F -\lambda/2 < \abs{k} < k_F + \lambda/2$. Then one divides this thin shell into $N$ patches with dimensionality $\sim \Lambda^{d-1} \times \lambda$ as shown in Fig.(2) in such a way that $\lambda \ll \Lambda \ll k_F$ and $\Lambda^2/k_F \ll \lambda$, where $d=2,3$ is the space dimension, $\Lambda$ is the linear dimension of the tangential extent of each patch. The condition $\lambda \ll \Lambda$ minimizes inter-patch scattering; $\Lambda \ll k_F$ and $\Lambda^2/k_F \ll \lambda$ together makes the curvature of the Fermi surface negligible. In the end we shall take the limit $\Lambda/k_F \rightarrow 0$, so that the sum over all patches can be converted to an integral over the Fermi surface. In this work, we treat the free theory in general $d$ dimensions, but shall restrict ourselves to $d=2$ when interactions are included.
\begin{figure}
\label{fig:patching} 
  \centering
  \subfigure[]{
    \label{fig:fs_devision} 
    \includegraphics[width=7cm]{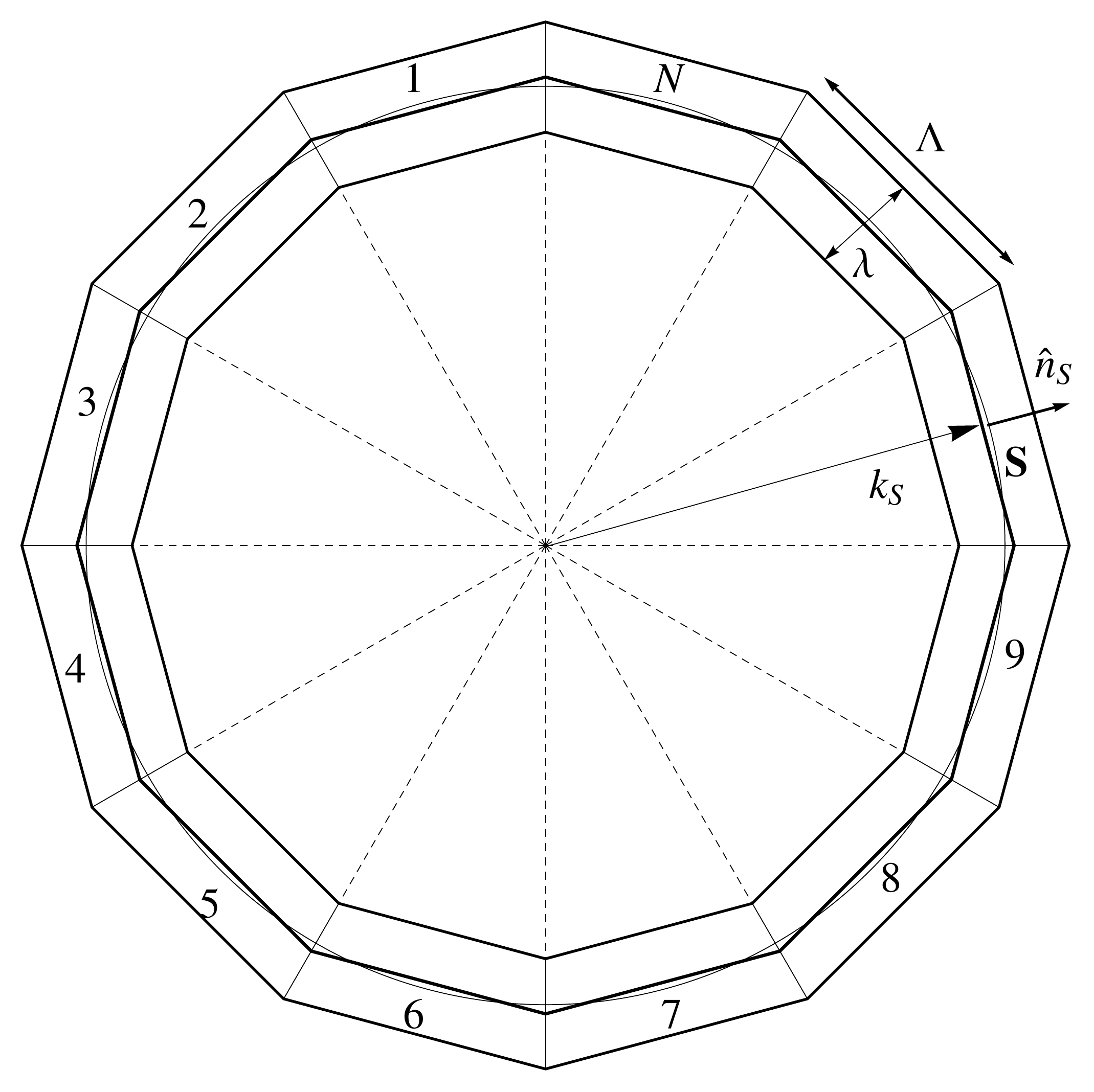}}
  \subfigure[]{
    \label{fig:patch} 
    \includegraphics[width=7cm]{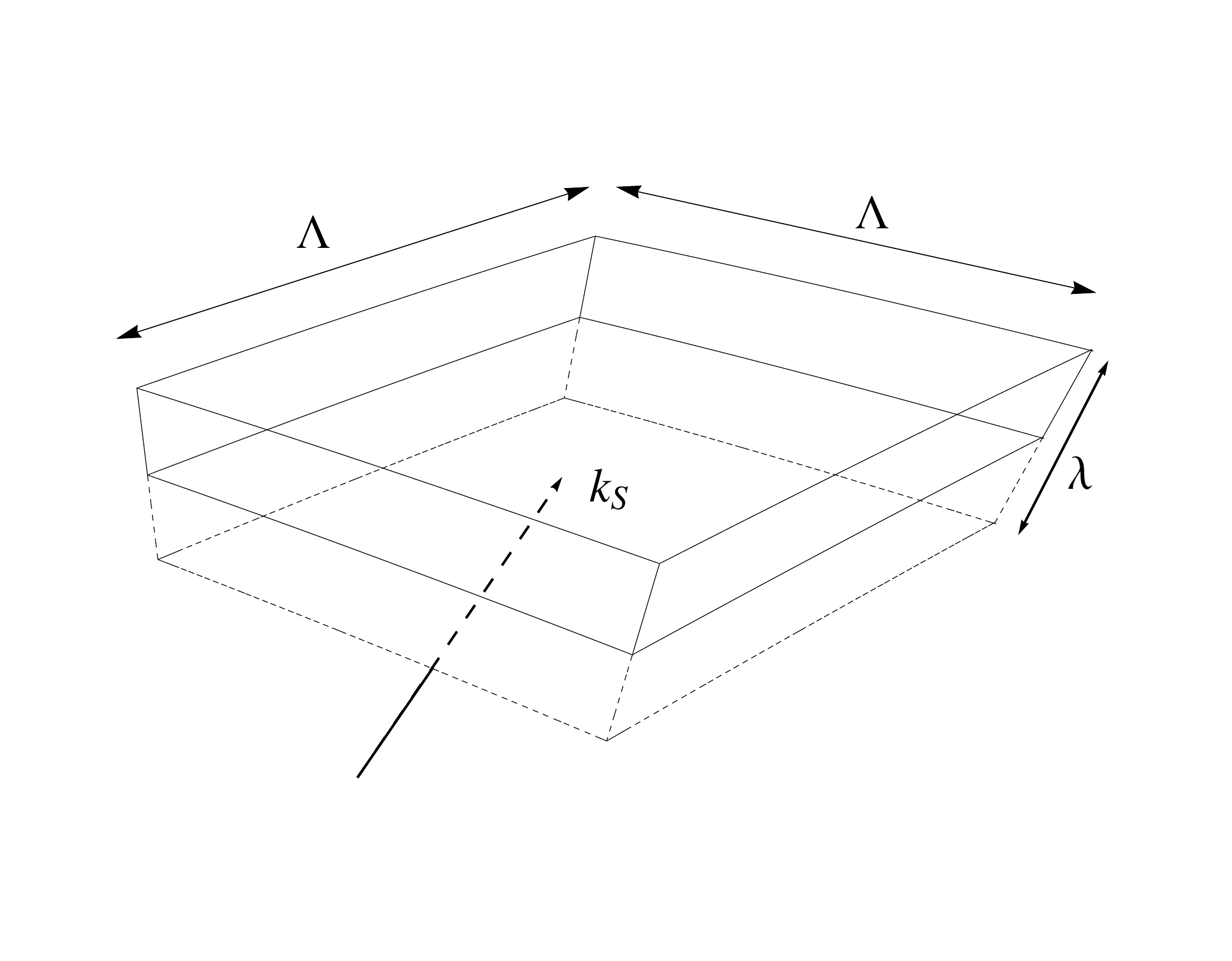}}
\caption{Patching of the Fermi surface. The low energy theory is restricted to within a thin shell about the Fermi surface with a thickness $\lambda \ll k_F$, in the sense of renormalization. The thin shell is further divided into $N$ different patches; each has a transverse dimension $\Lambda^{d-1}$ where $d=2,\ 3$ is the space dimensions. The dimensions of the patch satisfy three conditions: (1) $\lambda \ll \Lambda$ minimizes inter-patch scattering; (2) $\Lambda \ll k_F$ and $\Lambda^2/k_F \ll \lambda$ together makes the curvature of the Fermi surface negligible. (a): Division of a 2D Fermi surface into $N$ patches. Patch $\S$ is characterized by the Fermi momentum $\k_{\S}$. (b): A patch for $d=3$. The patch has a thickness $\lambda$ along the normal direction and a width $\Lambda$ along the transverse direction(s).}
\end{figure}
For an arbitrary patch $\S$, labeled by the Fermi momentum $\k_{\S}$ at the center of the patch, we introduce the patch fermion field operator \beq \psi(\S;\x) = e^{i\k_{\S}\cdot\x} \sum_{\p} \theta(\S;\p) e^{i(\p-\k_{\S})\cdot\x} \psi_{\p}, \eeq
where $\psi_{\p}$ is the usual fermion field in momentum space,
$$ \theta(\S;\p) = \begin{cases} 1 & \text{if $\p$ lies in the
patch $\S$,} \\ 0 & \text{if $\p$ lies outside patch $\S$.} \end{cases} $$
The effective Fermi liquid Hamiltonian can be written as
\beq
\begin{split}
& H[\psi^\dagger, \psi] = \int d^dx \sum_{\S}
\psi^\dagger(\S;\x)(\frac{\k_{\S}}{m^*} \cdot \nabla) \psi(\S;\x) \\
& + \int d^dx d^dy \sum_{\S,\T}  V(\S,\T;\x-\y) \psi^\dagger(\S;\x) \psi(\S;\x)\\
& \times \psi^\dagger(\T;\y) \psi(\T;\y),
\end{split}
\label{eq:H}
\eeq
with $m^*$ being the effective mass, $V(\S,\T;\x-\y)$ the effective interaction in the forward scattering channels. Even though this model is restricted to special interactions of this form, forward scattering is known to be the only marginal interaction in RG analysis\cite{RevModPhys.66.129}. As the leading order contribution of the entanglement entropy is dominated by the low energy modes around the Fermi surface, it is sufficient to consider this model. Similar to the 1D case, the bosonic degrees of freedom are the density modes of the system, in this case defined within each patch of the Fermi surface:
\beq\label{eq:current_operator}
J(\S;\q) = \sum_{\k} \theta(\S;\k-\q)
\theta(\S;\k)\{\psi^\dagger_{\k-\q} \psi_{\k} -
\delta^d_{\q,0} \ev{\psi^\dagger_{\k} \psi_{\k}}\}.
\eeq
Though $\q$ is not explicitly bounded in the above definition of the patch density operator, its transverse components $\q_{\S\bot} = (q_{\S\bot}^{(1)}, \dots,q_{\S\bot}^{(\alpha)},\dots,q_{\S\bot}^{(d-1)})$ (those parallel to the Fermi surface) are limited $ q_{\S\bot}^{(\alpha)} \in (-\Lambda, \Lambda)$ due to the patch confinement. Their commutation relation is
\beqarr\label{eq:j_commutator_1}
&\begin{split}
& [J(\S;\q), J(\bm{T};\p)] \simeq \delta_{\S,\T} \delta_{\q+\p,0}^d \sum_k \theta(\S;\k) \\
&\times [\theta(\S;\k-\q) - \theta(\S;\k + \q)] n_{\k}\\
\end{split}\\
\label{eq:j_commutator}
&= \delta_{\S,\T} \delta_{\q+\p,0}^d \Omega \ (\n_{\S} \cdot \q) \theta_2(\q_{\S\bot}) +\mathcal{O}(\lambda/\Lambda),
\eeqarr
where
\beq
 \theta_2(\q_{\S\bot}) = \prod_{\alpha} (1 - q_{\S\bot}^{(\alpha)}/\Lambda),
\eeq
$\Omega = \Lambda^{d-1} \left[L_0 / (2\pi) \right]^d$, $n_{\k} = \ev{\psi_k^\dagger\psi_k}$ is the occupation number of state with momentum $\k$, $\n_{\S}$ is the outward normal direction of patch $\S$, $\q_{\S\bot}$ represents all other component(s) of $\q$ that are perpendicular to $\n_{\S}$, and $L_0$ is the linear dimension of the entire system. The appearance of $\delta_{\q+\p,0}$ is a result of momentum conservation. The calculation of the commutator is reduced to computing the difference of occupied states, i.e. the area difference below the Fermi surface, between the two $\theta$ functions $(\theta(\S; \k - \q) - \theta(\S; \k - \p))$ as indicated by Eq. (\ref{eq:j_commutator_1}). This is similar to 1D bosonization. If we consider both $\k$ and $\q$ to be 1D momenta, Eq. (\ref{eq:j_commutator_1}) would give us the 1D bosonization commutator. The 2D result Eq. (\ref{eq:j_commutator}) is similar, because the Fermi surface confined within the patch is essentially flat thus the dispersion is 1D. That leads to the $\n_{\S}\cdot \q$ dependence of the commutator as that of the 1D case, even for $\q_{\S\bot} \neq 0$. The difference is that, as illustrated in Fig. (3),  due to the patch confinement on the transverse direction(s), when $\q_{\S\bot}$ increases $\k \pm \q$ would increasingly find itself outside the patch thus not contributing to the commutator. According to Fig. (3), one can see that this gives rise to the factor $\theta_2(\q_{\S\bot})$, which  diminishes the commutator at large $\q_{\S\bot}$.
\begin{figure}\label{fig:commutator}
  \includegraphics[width=8cm]{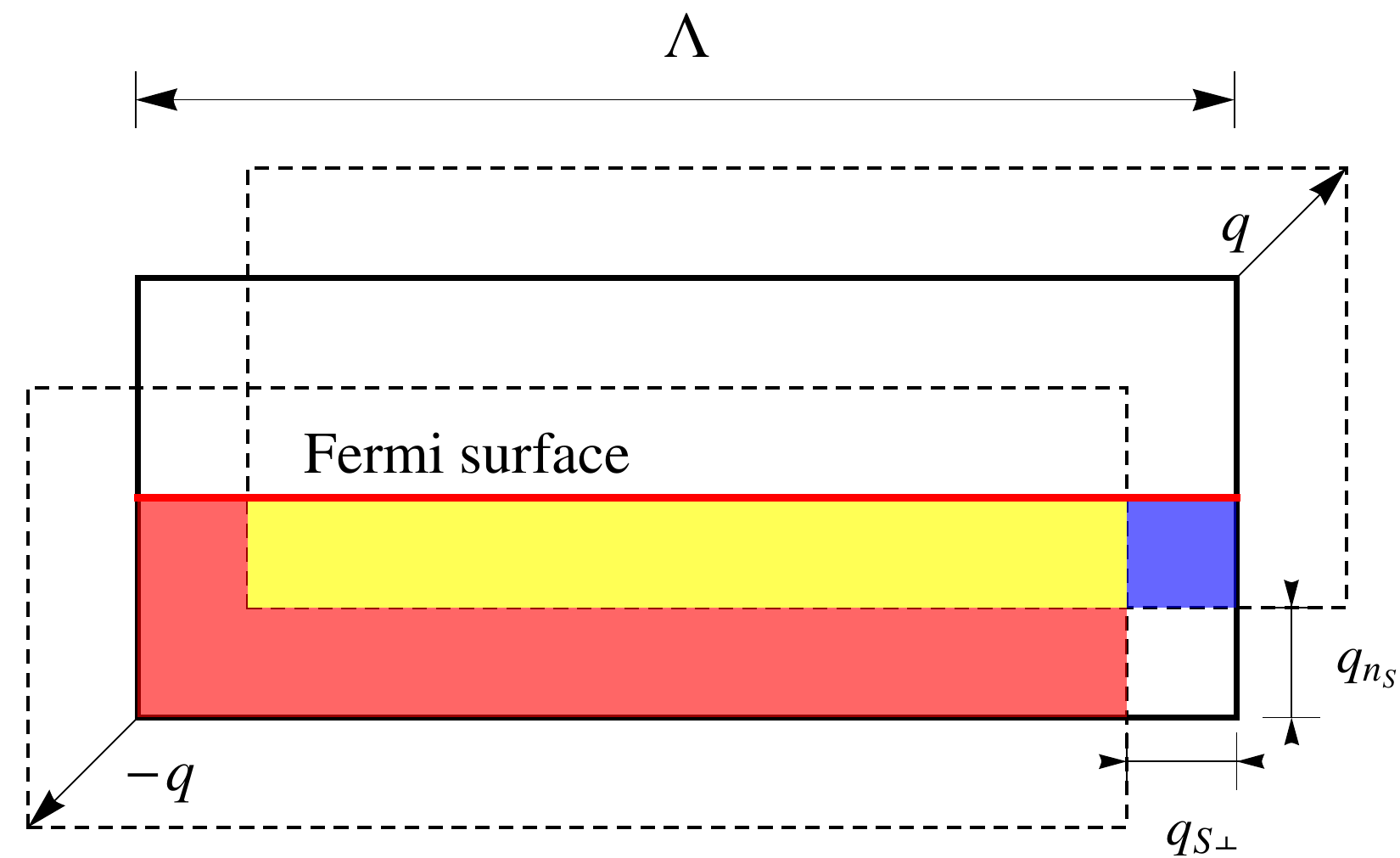}
  \caption{(Color online) Origin of the bosonic commutator of patch density operators illustrated for $d=2$. As shown in Eq. (\ref{eq:j_commutator}), the commutator is reduced to computing the difference of occupied states , i.e. the area difference below the Fermi surface, between the two $\theta$ functions $(\theta(\S; \k - \q) - \theta(\S; \k + \p))$. The solid box indicates the original patch, or $\theta(\S;k)$. The red line shows the Fermi surface. Both $\theta(\S; \k - \q) $ and $ \theta(\S; \k + \p)$ are denoted by dashed boxes. The occupied part in $\theta(\S; \k + \q)$ is denoted by blue,  that of $\theta(\S; \k - \q)$ is denoted by red, and the overlapping region is denoted by yellow. Subtracting the remaining blue area from the red, we obtain that $\theta(\S; \k - \q)$ occupies $(\Lambda - q_{\S\bot}) q_{\n_{\S}}$ more states, which gives us the commutator.}
\end{figure}
It is usually neglected in literature because the long wavelength limit is taken\cite{houghton1993, CastroNeto1994, CastroNeto1994a,  CastroNeto1995}. However, as it is important in the present context to correctly count the number of total degrees of freedom, this $\theta_2(\q_{\S\bot})$ factor cannot be neglected because it comes from counting the transverse degrees of freedom. To simplify things, we replace $\theta_2 (\q_{\S\bot})$ by
\beq
\theta_2(\q_{\S\bot}) = 1 \quad \text{for $-\Lambda/2 <  q_{\S\bot}^{(\alpha)} < \Lambda/2$,}
\eeq
and we also limit $\q_{\S\bot}$ to this range. This approximation makes it easier to do Fourier transform while keeping the total degrees of freedom intact. To see that, it is sufficient to consider one direction, comparing the area enclosed by the two different functions: $\theta_2(q_{\bot}) = 1-q_{\bot}/\Lambda$ over the range $(-\Lambda,\Lambda)$ and $\theta_2(q_{\bot})  = 1$ over the range $(-\Lambda/2,\Lambda/2)$. Both functions enclose the same area thus the same number of states. This approximation can also be interpreted as relaxation of the hard wall cutoff in Eq. (\ref{eq:current_operator}), softening of the step function $\theta(\S;\k)$. In Eq. (\ref{eq:current_operator}), $\q$ is not bounded while $\k$ is bounded by $\theta(\S;\k)$. If we relax the restriction on $\k$ on the transverse direction, allowing $\k$ with $\abs{k_{\S\bot}^{(\alpha)}} > \Lambda/2$ in the summation, but require $q_{\S\bot}^{\alpha}$ to be bounded within the patch, we would obtain the alternative $\theta_2(q_{\bot})$.

Using from now on the above approximation, we construct the local bosonic degrees of freedom $\phi(\S;\x) = \phi(\S; x_{\S}, \x_{\S\bot})$ as
\beq
J(\S;\x) = \sqrt{\Omega} \partial_{x_{\S}} \phi(\S; x_{\S},
\x_{\S\bot}),
\eeq
where $J(\S;\x) = \sum_{\q} e^{i\q\cdot\x} J(\S;\q)$, $ x_{\S} = \x \cdot \n_{\S}$, and $\x_{\S\bot} = \x  - (\x\cdot \n_{\S}) \n_{\S}$. The commutation relations for the $\phi$'s are then
\beq\label{eq:bcom}
\begin{split}
& [\partial_{x_{\S}} \phi(\S;\x), \phi(\T; \y)] =
i 2\pi\Omega\delta_{\S,\T} \delta(x_{\S} - y_{\S})\\
& \times \prod_{\alpha=1}^{d-1}\left( \frac{\sin(\Lambda (x_{\S\bot}^{(\alpha)} - y_{\S\bot}^{(\alpha)}))}{2\pi (x_{\S\bot}^{(\alpha)} - y_{\S\bot}^{(\alpha)})} \right) \\
\end{split}
\eeq
which is the bosonic commutation relation we are looking for. The factor $\prod_{\alpha}\left( \frac{\sin(\Lambda (x_{\S\bot}^{(\alpha)} - y_{\S\bot}^{(\alpha)}))}{2 \pi (x_{\S\bot}^{(\alpha)} - y_{\S\bot}^{(\alpha)})} \right)$ arising from transverse directions must be treated with care in different circumstances. In most literature, the focus is the physics at large length scale $l \gg 1/\Lambda$; therefore, this factor is usually approximated by $\delta^{d-1} (\x_{\S\bot} - \y_{\S\bot})$ which is good in that limit without further discussion. This is also what we shall do for most of the time unless noted otherwise:
\beq\begin{split}
& [\partial_{x_{\S}} \phi(\S;\x), \phi(\T; \y)]  \underrightarrow{^{\abs{x_{\S\bot}^{(\alpha)} - y_{\S\bot}^{(\alpha)}} \gg 1/\Lambda}} \\
&\simeq i 2\pi\Omega\delta_{\S,\T}^{d-1} \delta(x_{\S} - y_{\S}) \delta^{d-1} (\x_{\S\bot} - \y_{\S\bot}).
\end{split}
\eeq
However, the more accurate expression \eqref{eq:bcom} is useful for us to understand how to count the transverse degrees of freedom correctly. It tells us that the transverse degrees are not independent on the short length scale $l < 1/\Lambda$. More importantly, later on we need to consider the limit  $\delta^{d-1} (\x_{\S\bot} - \y_{\S\bot}) \vert_{\y_{\S\bot} \rightarrow \x_{\S\bot}}$; without Eq. (\ref{eq:bcom}), this limit would be ill-defined.

With the above, the Hamiltonian $H[\psi^\dagger, \psi]$ is found to be quadratic in terms of these $J(\S;\q)$'s:
\beq\label{eq:bosonized_h}
\begin{split}
H[\psi^\dagger, \psi] = \frac{1}{2} \sum_{\S,\T;\q} \frac{v_F^*\delta_{\S,\T}}{\Omega}J(\S;-\q) J(\T;\q) \\+ V(\S,\T; q) J(\S;-\q) J(\T;\q),
\end{split}\eeq
where $V(\S,\T;\q)$ is the Fourier transform of $V(\S,\T; \x-\y)$. So it is also quadratic in the bosonic fields associated with the $J(\S;\q)$'s.

\subsection{Entanglement Entropy of Free Fermions}
 The kinetic energy part of Eq. (\ref{eq:H}) or its bosonized version Eq. (\ref{eq:bosonized_h}) can be written in terms of the boson fields constructed above as:
\beq\label{eq:bH_free}\begin{split}
H_0 &= \frac{1}{2} \sum_{\S;\q} \frac{v_F^*}{\Omega}J(\S;-\q) J(\S;\q)\\
&= \frac{2\pi v_F^{*}}{\Omega V} \sum_{\S} \int
d^2x \left(\partial_{x_{\S}} \phi(\S;\x)\right)^2.
\end{split}\eeq
We see that there is no coupling between different patches. The theory is thus formally a tensor product of many independent theories, one for each patch. We can therefore calculate the entanglement entropy patch by patch and sum up contributions from each patch in the end. Within a single patch there is no dynamics in the perpendicular direction as dictated by the Hamiltonian, and the problem is reduced to a {\it one dimensional} problem! Note that transverse degrees of freedom are not completely independent. According to Eq. (\ref{eq:bcom}), the commutator is non-vanishing for $x_{\S\bot} \neq y_{\S\bot}$ up to a length scale $\sim 2\pi/\Lambda$. This is a consequence of restricting $\q_{\S\bot}$ to within the range $[-\Lambda/2,\Lambda/2]$. Physically one can view this as {\it discretization} along the transverse direction due to a restricted momentum range, similar to the relation between a lattice and its Brillouin zone. In this view the single patch problem is reduced to a 1D problem with a chain density of  $\left(\Lambda/(2\pi)\right)^{d-1}$. Therefore, the Hamiltonian (\ref{eq:bH_free}) becomes
\beq
H_0 = \frac{2\pi v_F^{*}}{\Omega V} \sum_{\S; \x_{\S \bot}} \int
dx_{\S} \left(\partial_{x_{\S}} \phi(\S;\x)\right)^2.
\eeq

Note that the bosonized theory of a single patch is chiral. To directly make use of our toy model, we need to consider two patches having opposite $\n_{\S}$ simultaneously. This is because for a 1D fermion model at non-zero filling, there are two Fermi points. Both need to be considered to construct well-defined local degrees of freedom. Once we consider such two patches together, it is more convenient to combine the two chiral theories into a non-chiral theories. This is also what we will do for the rest of this work. Introduce the non-chiral fields
\beq
\begin{cases}
\varphi(\bm{S};\bm{x}) = \frac{1}{\sqrt{2}} \left(
\phi(\bm{S};\bm{x}) - \phi(-\bm{S};\bm{x})\right), \\
\chi(\bm{S};\bm{x}) = \frac{1}{\sqrt{2}} \left(
\phi(\bm{S};\bm{x}) + \phi(-\bm{S};\bm{x})\right),
\end{cases}
\eeq
where $-\S$ indicates the patch with normal direction opposite to that of patch $\S$: $\n_{-\S} = - \n_{\S}$. One finds that $\chi$ and $\varphi$ are
mutually dual fields
with $\S$ restricted to one hemisphere, but $\partial_{x_{\S}}\varphi$ and $\varphi$ now commute while $\chi$ and $\varphi$ have a non-trivial commutator:
 \beq\begin{split}
 &[\varphi(\S;\x),\partial_{y_{\S}}\varphi(\S;\y)] =[\chi(\S;\x),\partial_{y_{\S}}\chi(\S;\y)] = 0, \\
 &[\partial_{x_{\S}}\varphi(\S;\x), \chi(\T;\y)] = [\partial_{x_{\S}}\chi(\S;\x), \varphi(\T;\y)] \\
&= 2 i \pi \Omega \delta_{\S,\bm{T}} \delta(x_{\S} - y_{\S}) \delta^{d-1}(\x_{\S\bot} - \y_{\S\bot}).
\end{split}
 \eeq
Therefore, two patches with opposite $\n_{\S}$ are equivalent to a set of ordinary 1D boson fields. Throughout the rest of this work, we shall assume this chiral-to-nonchiral transformation is done, and when we refer to patches we always refer to the two companion patches that form a non-chiral patch together. For the non-chiral boson theory, it is known that the entanglement entropy of a single interval (with two end points) is $(1/3) \log L$.

Before we proceed further, we note that the relation between boson fields and the original fermion fields is not completely local. However, the underlying physical quantity that matters is not the fields, but the fermion density, or in other words, the fermion number basis one chooses to expand the Hilbert space of the problem. This physical basis is also what one uses to do the partial trace. It is known that the fermion density operator obeys a locally one-to-one corresponding relation to the boson fields. Thus we argue that in 1D the nonlocal relation between the fermion and boson fields does not affect the partial trace operation, so as the calculation of entanglement entropy.

By referring to our result for the toy model, the contribution from a single patch is readily given
\beq
S(\S) = \frac{1}{12}\log L \oint_{\partial A} \abs{\n_{\S} \cdot
  d\vec{S}_x} \times \left(\frac{\Lambda}{2\pi}\right)^{d-1},
\eeq
where an additional factor of $1/2$ has been introduced in order to count only once each pair of patches forming a non-chiral theory.
Identifying $\n_{\S} \Lambda^{d-1}$ as the surface element at the Fermi surface $d\vec{S}_k$ and taking the $N \rightarrow \infty$ limit, the total entanglement entropy is
\beq
\begin{split}\label{eq:GK}
S 
& = \frac{1}{12(2\pi)^{d-1}} \log L \oint_{\partial A} \oint_{\partial
\Gamma} \abs{d\vec{S}_k \cdot d\vec{S}_x}.
\end{split}
\eeq
So we recover the GK result for generic free fermions.

\subsection{Solution for the Fermi liquid case and non-locality of the Bogoliubov fields}
When Fermi liquid interactions (forward scattering) are included, the full Hamiltonian will no longer be diagonal in the patch index $\S$. But it is still quadratic in terms of the patch density operators, i.e. the bosonic degrees of freedom, and can be diagonalized by a Bogoliubov transformation. According to Eq. (\ref{eq:j_commutator}) and ignoring terms of $\mathcal{O}(\lambda/\Lambda)$, one can define a set of boson creation/annihilation operators $\a^\dagger(\q) / \a(\q)$ as follows:
\begin{equation}\label{eq:cna_operator}
  \phi(\S;\x) = i \sum_{\q, \n_{\S}\cdot \q > 0} \frac{a^\dagger(\S;\q) e^{-i \q \cdot \x} - a(\S;\q) e^{i\q\cdot \x}}{\sqrt{\abs{\n_{\S} \cdot \q}}  }.
\end{equation}

It can be shown that the full Hamiltonian is diagonal in $\q$, and it can be diagonalized by a Bogoliubov transformation\cite{CastroNeto1995} independently for each $\q$ sector. In Ref. [\onlinecite{CastroNeto1995}], only a Hubbard-$U$ like interaction is considered for practical reasons. But in principle, such a Bogoliubov transformation also applies to general interactions:
\beq\begin{split}
&\a_i(\q) = \sum_{j} u_{ij} \alpha_j(\q) + v_{ij} \beta^\dagger_j(\q),\\
&\b_i(\q) = \sum_{j} u_{ij} \beta_j(\q) + v_{ij}\alpha^\dagger_j(\q),
\end{split}
\eeq
where both $i$ and $j$ refer to the patch index, $\alpha_j$ and $\beta_j$ are the Bogoliubov bosonic annihilation operators that diagonalize the Hamiltonian. With proper choice of $u$'s and $v$'s, the Hamiltonian is readily diagonalized. Ref. [\onlinecite{CastroNeto1995}] solves the Hubbard-$U$ like interaction and provides a successful description of Fermi liquids, even in the strong $U$ limit.

However, even for this simple case in which $U$ has no $\q$-dependence, the Bogoliubov transformation still depends on $\q$. To be more precise, $u_{ij}$ and $v_{ij}$ will depend
only on the angle between the patch normal direction $\n_{\S}$ and $\q$, leading to discontinuities in the derivatives at $q=0$.
 Consequently, the real space fields constructed from the Bogoliubov operators $\alpha_j$ and $\beta_j$ are {\it no longer} local with respect to the original
 boson fields.
 The real space Bogoliubov fields are constructed in a manner similar to Eq. (\ref{eq:cna_operator}):
\beq
\tilde{\phi}(\S;\x) = i \sum_{\q, \n_{\S} \cdot \q>0}
\frac{\alpha^\dagger(\S;\q) e^{-i\q\cdot\x} - \alpha(\S;\q)
  e^{i\q\cdot\x}}{\sqrt{\n_{\S} \cdot \q}}.
\eeq
Then one can show that  the original local degrees of freedom $\phi(\S;\x)$ can be expressed in terms of above Bogoliubov fields as
\beq\label{eq:nonlocality}
\begin{split}
\phi(\S;\x) &= \tilde{\phi}(\S;\x) + \int d\y \sum_l f(\S,l;\x-\y) \tilde{\phi}(l;\y),
\end{split}
\eeq
where $f(\S,l;\x - \y)$ is typically long-range, even for the short-range Hubbard-$U$ interaction. For more general cases, with further $\q$-dependence in the interaction, the non-locality would only be enhanced. The loss of locality prevents us from calculating the entanglement entropy directly using those eigen modes, since it is difficult to implement the partial trace using those non-local degrees of freedoms.
Therefore, although the Bogoliubov fields have a local core as we would expect for Fermi liquids from adiabaticity, they {\it do} acquire a nonlocal dressing due to interaction. Though in principle the partial trace can be done with those Bogoliubov fields, such nonlocality makes it difficult and we have not been able to do it, which further renders calculating the entanglement entropy impossible. This is very different from the 1D theory, where  for local interactions the eigen fields remain local, since there are only two Fermi points. There the transformation can never involve such angular $\q$-dependence due to limited dimensionality. Despite these technical difficulties, the non-locality may suggest possible corrections to the entanglement entropy. This is indeed the case as revealed by our later calculation for Fermi liquid interactions, although in this case such extra contributions are only of $\mathcal{O}(1) \times \log L$ which is of $\mathcal{O}(1/L)$ comparing to the leading term. This shows that the mode-counting argument in Ref.[\onlinecite{Swingle2010}], though correctly suggesting the $\log L$ violation to the area law for Fermi liquids, does not always fully account for all sources of entanglement entropy.

\section{Entanglement entropy from the Green's function}\label{sec:ee}
In order to preserve locality, we need to work with the original local  degrees of freedom. To do that, we adopt the approach used by Calabrese and Cardy\cite{calabrese2004} (CC) on calculating the entanglement entropy of a  free massive $1D$ bosonic field theory. The calculation is done in terms of the Green's function by applying the replica trick. In our case, we find that the CC approach can be generalized in a special geometry for solving the interacting theory which is quadratic after bosonization. In this way, we avoid diagonalizing the Hamiltonian and thus the nonlocality issue. However, we do have to regularize the theory by adding a mass term by hand. In the end we shall take the small mass limit, and replace the divergent
correlation
length $\xi \sim 1/m$ by the subsystem size $L$. The regularization procedure facilitates the calculation, but also strictly restricts us to computing the entanglement entropy only at the $\log L$ level.

In this section, by using the replica trick we convey the calculation of entanglement entropy into computing the Green's function on an $n$-sheeted replica manifold. We first demonstrate the method by applying it to free fermion theory in $d$-dimensional space; then based on it we compute the entanglement entropy perturbatively for a simple Fermi liquid theory in powers of the interaction strength up to the second order.

\subsection{The Replica Trick and Application to 1D Free Bosonic Theory}\label{sec:ee_free_fermion}
In this part, we briefly describe the replica trick in $(1+1)$ space-time dimensions ($(1+1)d$) so that later on we can straightforwardly generalize it to $(2+1)$ space-time dimensions ($(2+1)d$) accordingly for our problem.

\begin{figure}
  \centering
  \includegraphics[width=8cm]{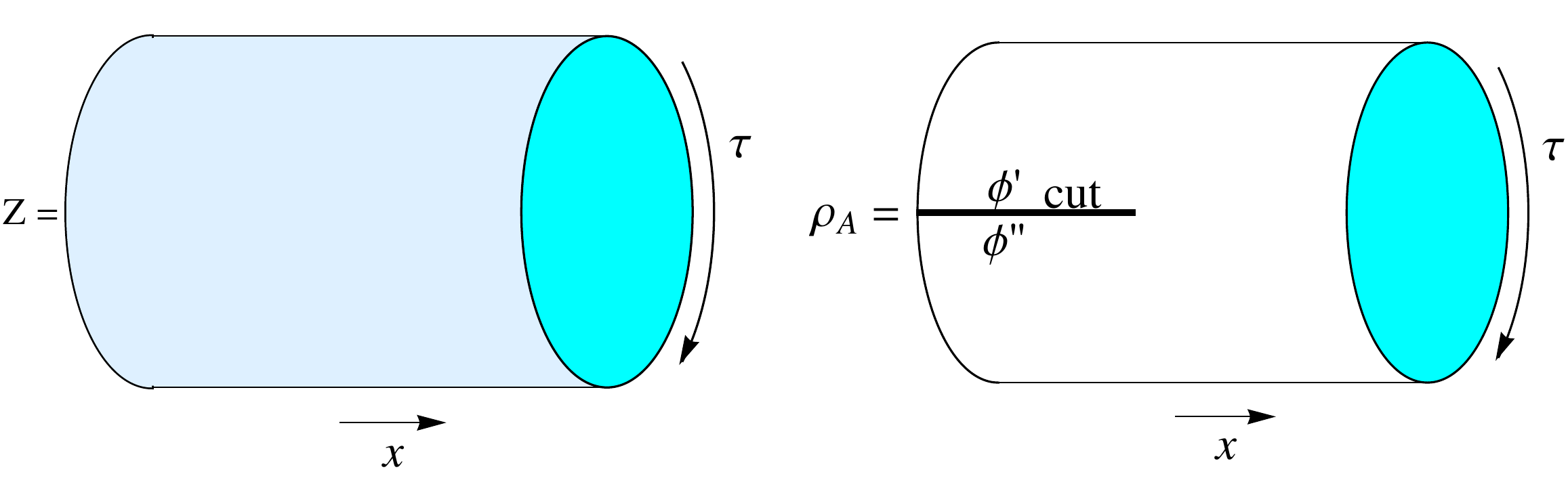}
  \caption{(Color online) Path integral representation of the reduced density matrix. Left: When we sew $\phi(x)' = \phi(x)'{'}$ together for all $x$'s, we get the partition function $Z$. Right: When only sew $x \not\in A$ together, we get $\rho_A$.}
  \label{fig:rdm}
\end{figure}

The replica trick makes use of the following identity:
\beq
S_A = -\tr (\rho_A \ln \rho_A) = -\lim_{n \rightarrow 1}
\frac{\partial}{\partial n} \tr \rho_A^n.
\eeq\label{eq:replica}To compute  $\tr \rho_A^n$,  CC use path integral to express the density matrix $\rho$ in terms of the boson fields
\beq
\rho(\{\phi(x)\}|\{\phi(x')'\}) = Z^{-1} \meanvalue{\{\phi(x)\}}{e^{-H}}{\{\phi(x')'\}},
\eeq
where $Z = \tr e^{-\beta H}$ is the partition function, $\beta$ is the inverse temperature, and $\{\phi(x)\}$ are the corresponding eigenstates of $\hat{\phi}(x)$: $\hat{\phi}(x) \ket{\{\phi(x')\}} = \phi(x') \ket{\{\phi(x')\}} $. $\rho$ can be expressed as a (Euclidean) path integral:
\beq\begin{split}
&\rho = Z^{-1} \int [d \phi(x,\tau)] \prod_x\delta(\phi(x,0)-\phi(x)')\\
& \times \prod_x\delta(\phi(x,\beta) - \phi(x)'') e^{-S_E},
\end{split}
\eeq
where $S_E = \int^\beta_0 L_E d\tau$, with $L_E$ being the Euclidean Lagrangian. The normalization factor $Z$, i.e. the partition function is found by setting $\{\phi(x)''\} = \{\phi(x)' \}$ and integrating over these variables. This has the effect of sewing together the edges along $\tau = 0$ and $\tau = \beta$ to form a cylinder of circumference $\beta$ as illustrated in Fig. (\ref{fig:rdm}) (left panel).

\begin{figure}
\label{fig:replica} 
  \centering
  \subfigure[]{
    \label{fig:n_rdm} 
    \includegraphics[width=7cm]{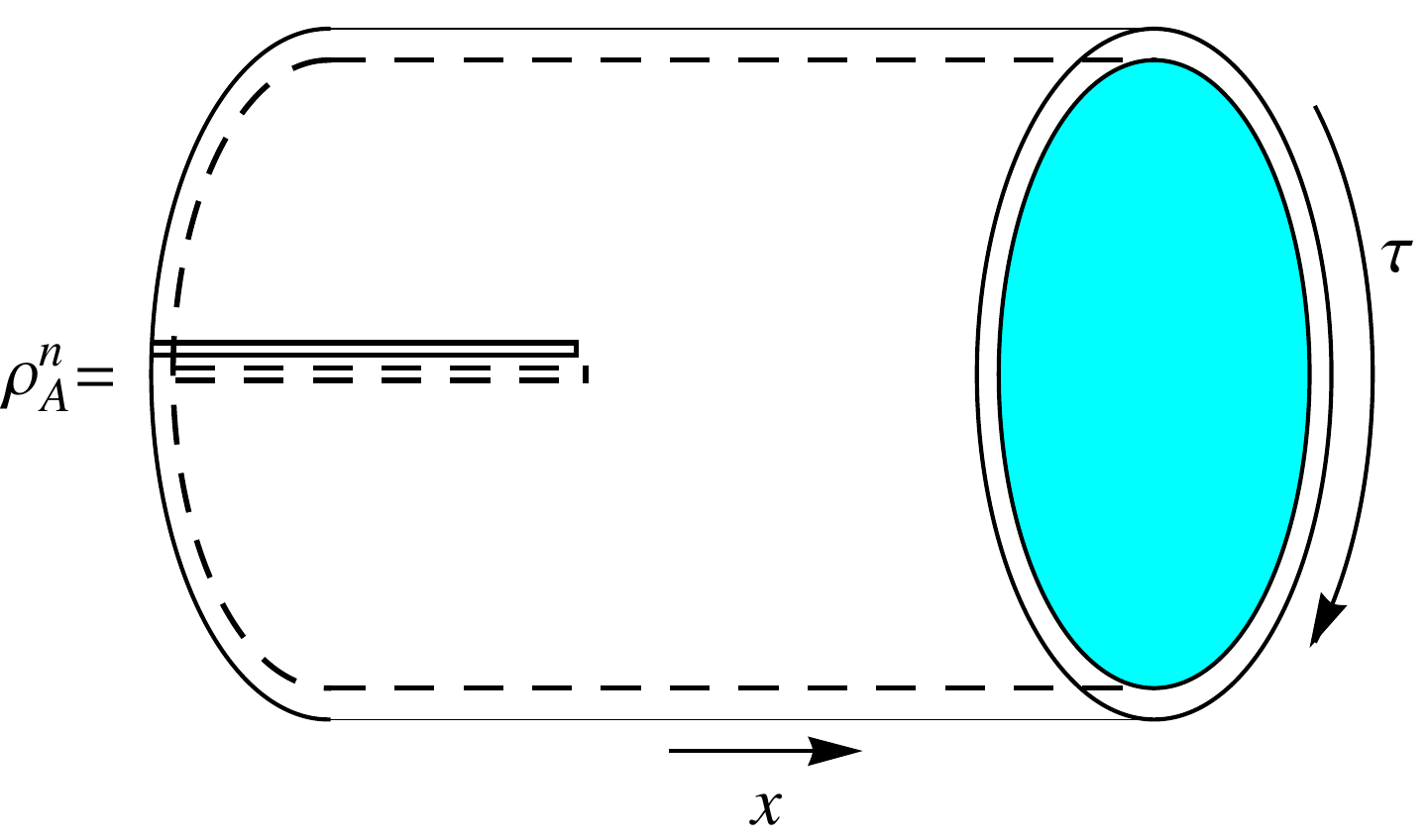}}
    \hspace{1.5cm}
  \subfigure[]{
    \label{fig:riemann_surface} 
    \includegraphics[width=7cm]{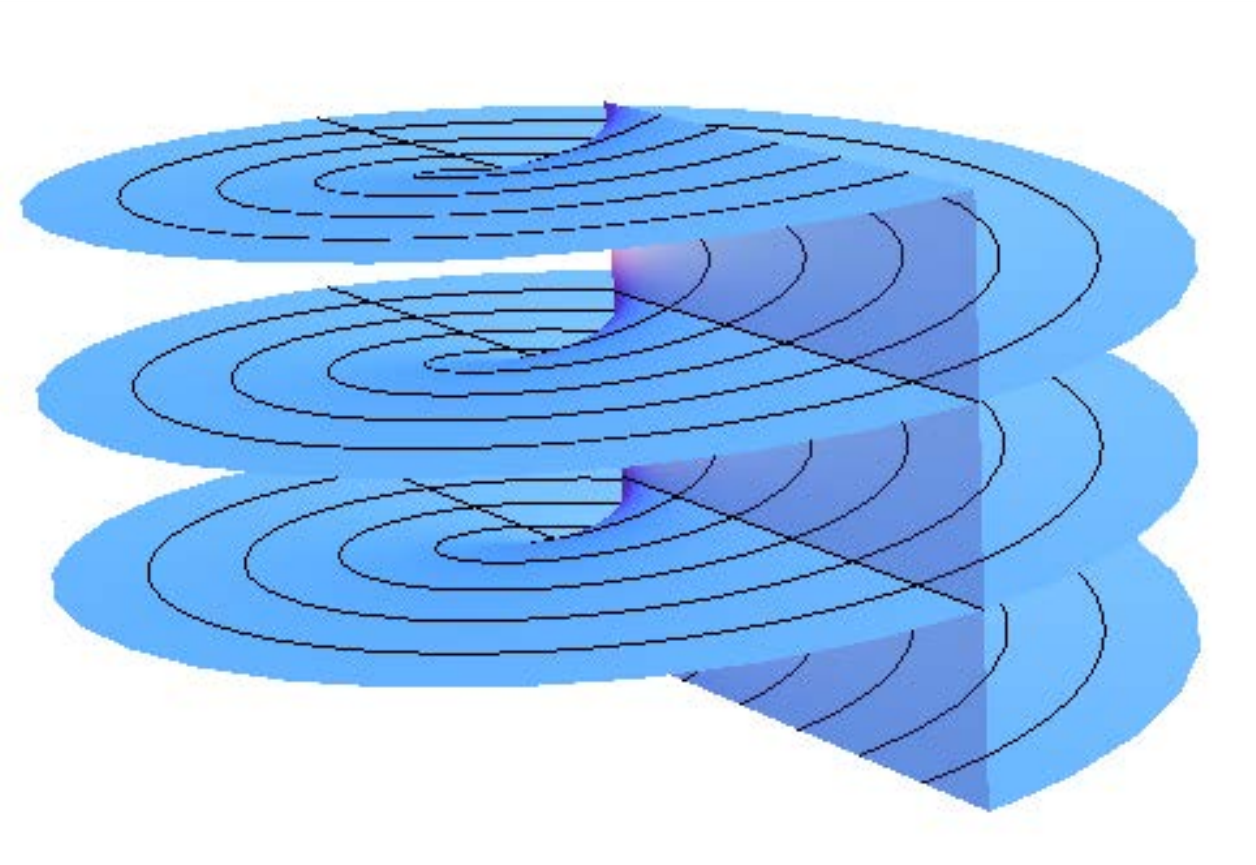}}
\caption{(Color online) Formation of the $n-$sheeted Riemann surface in the replica trick. By sewing $n$ copies of the reduced density matrices together, one obtains the replica partition function $Z_n$. In the zero temperature limit, $\beta \rightarrow \infty$, each cylinder representing one copy of $\rho_A$ becomes an infinite plane. Those $n-$planes sewed together form a $n-$sheeted Riemann surface in Fig. \ref{fig:riemann_surface} which can be simply realized by enforcing a $2 n \pi$ periodicity on the angular variable of the polar coordinates of the $(1+1)d$ plane instead of the usual $2\pi$ one. (a): $n$ copies of the reduced density matrices. For clarity only $n=2$ is shown. (b): Visualization of a $n-$sheeted Riemann surface.}
\end{figure}

The reduced density matrix of an interval $A = (x_i,x_f)$ can be obtained by sewing together only those points which are not in the interval $A$. This has the effect of leaving an open cut along the line $\tau = 0$ which is shown in Fig. \ref{fig:rdm} (right panel). To compute $\rho_A^n$, we make $n$ copies of above set-up labeled by an integer $k$ with $1 \le k \le n$, and sew them together cyclically along the open cut so that $\phi(x)'_k = \phi(x)''_{k+1} [\text{and }\phi(x)'_n = \phi(x)''_1]$ for all $x \in A$. In Fig. \ref{fig:n_rdm} we show the case $n=2$. Let us denote the path integral on this $n$-sheeted structure (known as $n$-sheeted Riemann surface) by $Z_n(A)$. Then
\beq
\tr \rho_A^n = \frac{Z_n(A)}{Z^n},
\eeq
so that
\beq
S_A = -\lim_{n \rightarrow 1} \frac{\partial}{\partial n}
\frac{Z_n(A)}{Z^n}.
\eeq

If we consider the theory as that of one field living on this complex $n-$sheeted Riemann surface instead of a theory of $n$ copies, it is possible to remove the replica index $n$ from the fields, and instead consider a problem defined on such an $n$-sheeted Riemann surface which can be realized by imposing proper boundary conditions.

In Ref.[\onlinecite{calabrese2004}], CC consider the entanglement entropy between the two semi-infinite 1D system (i.e. cutting an infinite chain into two halves at $x=0$) for free massive boson fields.
For such geometry, as illustrated in Fig. \ref{fig:riemann_surface}, the $n-$sheeted Riemann surface constraint is realized by imposing a $2 n \pi$ periodicity on the angular variable of the polar coordinates of the $(1+1)d$ plane instead of the usual $2\pi$ one. In this way, the $(1+1)d$ variable $\x=(x,\tau)$ acquires $n$ branches $\x_{n}$, and each branch corresponds to one copy of $\phi$. Notation-wise this corresponds to
\beq
\phi(x,\tau)_k \Rightarrow \phi(\x_k) \Rightarrow \phi(\x),
\eeq
and the sewing conditions $\phi(x)'_k = \phi(x)''_{k+1}$ simply becomes the continuity condition for $\phi(\x)$ across its consecutive branches. Here we use a generalized polar coordinate: $\x = (r,\theta)$ with $0< r < \infty$, and $0 \le \theta < 2n \pi$.

The massive free boson theory considered by CC is defined by the following action
$$\mathcal{S} = \int \frac{1}{2}((\partial_{\mu} \phi)^2 - m^2 \phi^2)d^2r.$$
The $(1+1)d$ bosonic Green's function $G_{0,b}^{(n)}(\r,\r') = \ev{\phi(\r)\phi(\r')}$ on the $n-$sheeted Riemann surface satisfies the differential equation
$$(-\nabla^2_{\r} +m^2)G_{0,b}^{(n)} = \delta(\r-\r').$$
To compute the partition function, one can make use of the identity
\beq
\frac{\partial}{\partial m^2} \log Z_n = -\frac{1}{2} \int d^{d+1} x G^{(n)}(\x,\x).
\eeq
Note that here the integration is over the entire $n-$sheeted space. 
The above is applicable to general quadratic theories of bosons, and will be applied by us later to bosonized theories of interacting fermions.
Here we use $G^{(n)}(\x,\x')$, a general two point correlation function on the $n-$sheeted Riemann surface in $d$-dimensional space for later use,
instead of the specific $G_{0,b}^{(n)}$ defined above. Accordingly, $S_A$ is then given as
\beq\label{eq:entropy}
S_A =  - \lim_{n \rightarrow 1} \frac{\partial}{\partial n} e^{- \frac{1}{2} \int dm^2  \int d^{D+1} x (G^{(n)}(\x,\x) - n G^{(1)}(\x,\x))}.
\eeq
Here and in the following, we will leave it understood that the first term in the integrand is integrated over the $n$-sheeted geometry,
whereas the second is integrated over a one-sheeted geometry. There should be no confusion as the superscript of $G$ generally
indicates the geometry.

The benefit of the above approach is that the two point correlation function or Green's function, defined in terms of certain differential equation obtained from the equation of motion,
can be solved for on the $n-$sheeted Riemann surface thus enabling us to compute the entanglement entropy.
Although CC's work only considers massive $(1+1)d$ boson fields, it is also applicable to our case. The price one has to pay is to to introduce a mass term for regularization. At the end of the calculation the inverse mass, which is the
correlation
length of the system, shall be considered to be on the same scale as $L$: $1/m \sim L$, where $L$ is the characteristic length scale of the subsystem. The validity of such
consideration is well-established in other cases,\cite{Korepin2004,Hertzberg2011} where the
correlation
 length is either set by finite temperature or mass. The only modification necessary to apply the above to
 a bosonized Fermi surface in higher dimensions
  is to introduce a sum over the patch index.


\subsection{Geometry and Replica Boundary Conditions}
\begin{figure}
 \centering
  \subfigure[]{
   \label{subfig:geometry}
  \includegraphics[width=8cm]{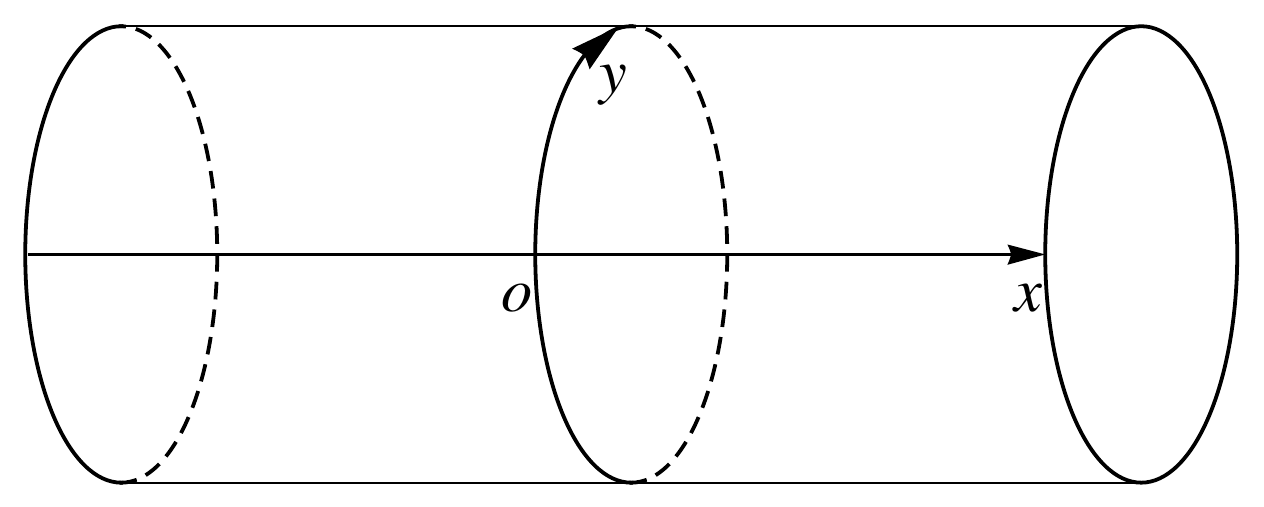}}
\hspace{1cm}
\subfigure[]{
    \label{subfig:bc} 
    \includegraphics[width=9cm]{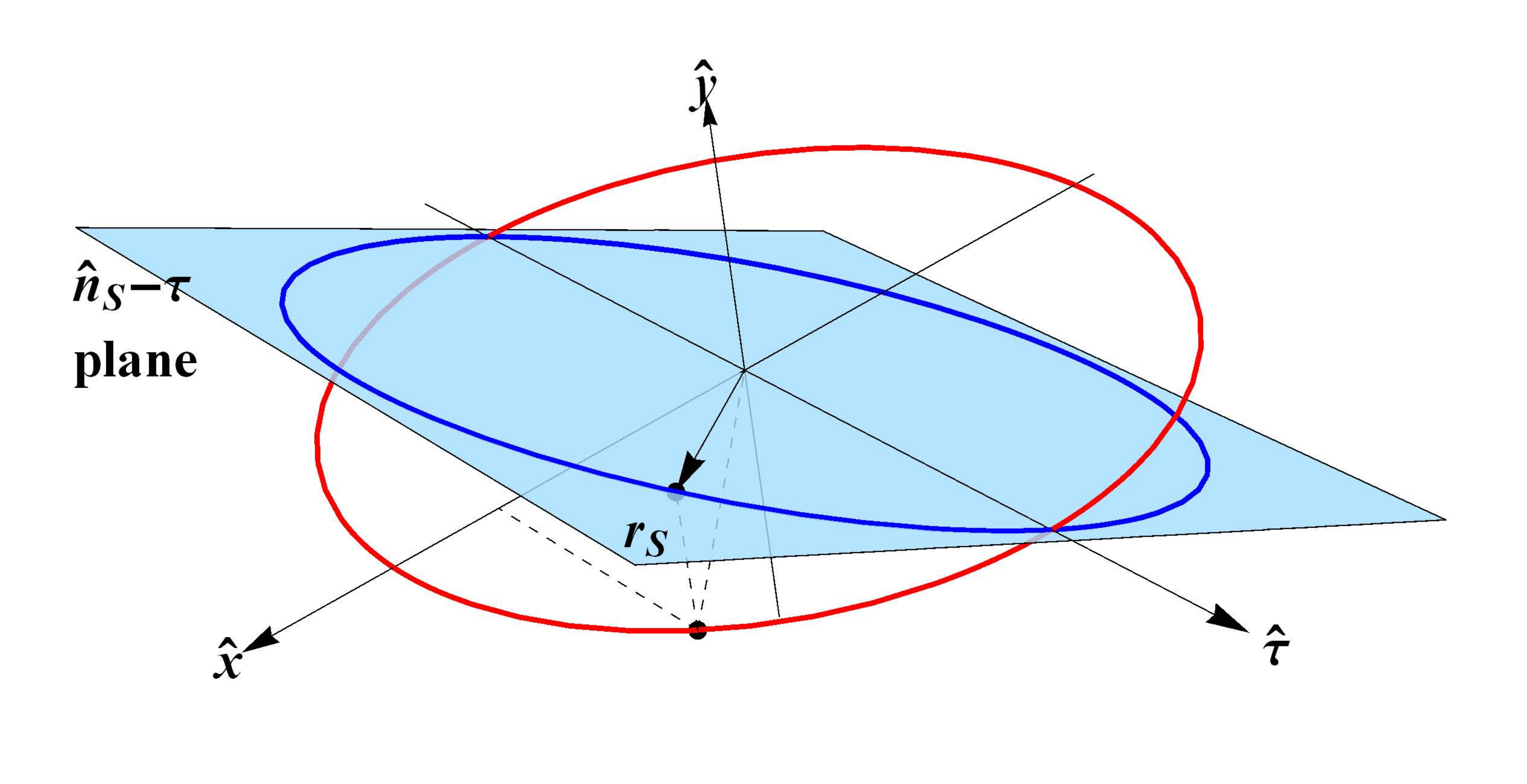}}
  \caption{(Color online) The half-cylinder geometry and equivalence of boundary conditions in $\hat{x}-\hat{\tau}$ and $\n_{\S}-\hat{\tau}$ planes. The system is infinite in the $\hat{x}$ direction while obeys periodic boundary condition along the $\hat{y}$ direction with length $L$. The system is cut along the $\hat{y}$ axis so that we are computing the entanglement entropy between the two half planes. (a): The half-cylinder geometry.  (b): The projection of $\r_{\S}$ onto the $\hat{x}-\hat{\tau}$ plane. Consider polar coordinates of an arbitrary $\n_{\S}-\hat{\tau}$ plane (the blue plane). Since the polar coordinates in the $\hat{x}-\hat{\tau}$ plane satisfies the $2n \pi$ periodic boundary condition, consider the one-to-one projection of the vector $\r_{\S}$ onto the $\hat{x}-\hat{\tau}$ plane. Consider, if we move the vector in the $\hat{x}-\hat{\tau}$ plane around the origin $n$ times (the red circle). Due to the one-to one mapping, $\r_{\S}$ should also move around the origin $n$ times (the blue "circle", it is actually a eclipse), thus obeys the $2n\pi$ periodicity as well.}
  \label{fig:geometry}
\end{figure}

Through the remainder of this work, instead of the general geometry considered before, we work with a special half-cylinder geometry as shown in Fig. \ref{subfig:geometry}: the system is infinite in the $\hat{x}$ direction while obeying periodic boundary condition along the $\hat{y}$ direction with length $L$. The system is cut along the $\hat{y}$ axis so that we are computing the entanglement entropy between the two half planes. We require $L$ to be large so that it can be considered $\sim \infty$ unless otherwise noted.

We choose such this simple geometry for the following reasons.
Cutting the system straight along the $\hat{y}$ direction, yielding a two half-plane geometry, is a straightforward $(2+1)d$ generalization of the semi-infinite chain geometry considered in CC. It makes any straight line intersect the boundary only once, dividing it into two  semi-infinite segments, for all patch directions as in the 1D case, except for lines parallel to the $\n_{\S} = \hat{y}$ patch direction.
The degrees of freedom associated with this special patch do not contribute to the entanglement entropy, since they are not coupled
 (have no dynamics) along $\hat{x}$, and are of measure zero in the large patch number limit anyway.

For this simple geometry, the $(2+1)d$ $n$-sheeted geometry is constructed from $n$ identical copies
\begin{equation}
    S^n=\{ (x,y,\tau) \in \mathbb{R}\times\mathbb{R}\times\mathbb{R}  \}\,,
\end{equation}
sliced along ``branch cuts''
\begin{equation}
    C^n=\{ (x,y,\tau) \in \mathbb{R}^- \times\mathbb{R}\times\{ 0\}  \}\,,
\end{equation}
and then appropriately glued together along these cuts. This happens exactly as in 1D, and the $y$ coordinate is so far a mere spectator. This defines an $n-$sheeted, or in this case more appropriately the $n-$layered, replica manifold which is a simple enough generalization of the $(1+1)d$ case. The $n$-sheeted Riemann surface, as discussed in Sec. \ref{sec:ee_free_fermion} and shown in Fig. \ref{fig:riemann_surface}, now acquires an extra direction $\hat{y}$ perpendicular to the $\hat{x} - \hat{\tau}$ plane.
It can still be implemented by imposing the same $2n\pi$ periodicity boundary conditions on $\theta$, the angular variable of the polar coordinates $(x,\tau) = (r \cos \theta, r \sin \theta)$ in the $\hat{x} - \hat{\tau}$ plane. Therefore, we can safely make use of the CC result, i.e. the solution to the Green's function on a $n-$sheeted Riemann surface, to the free fermion theory, and can further use it as a starting point for treating the interacting theory. This is obviously true for the patch with $\n_{\S} = \hat{x}$, but it also holds for general $\n_{\S}$ as we shall validate as the following.

For a general patch direction $\n_{\S}$, the noninteracting Green's function associated with this patch embodies correlations in the affine $\n_{\S} - \hat{\tau}$ ``planes''.
The geometry of each such ``plane'' is that of the $n$-sheeted Riemann surface of the $(1+1)d$ problem, as we will now argue.
With each patch direction we thus associate a different foliation of the $n$-sheeted $(2+1)d$ geometry into $(1+1)d$ counterparts.

To be more precise, for given patch $S$, instead of the Cartesian coordinates $(x,y,\tau)$, we consider a parallel/perpendicular decomposition $(x_{\S},x_{\S\perp},\tau)$ for each sheet
via
\begin{equation}\label{ct1}
   (x,y)=x_{\S}  \mathbf{\n}_{\S}+x_{\S\perp}\mathbf{\n}_{\S \perp} \,,
\end{equation}
where the $\n_{\S\perp}$ are the perpendicular unit vectors aligned with the patch $S$.
The natural choice of coordinates for a given patch is to choose polar coordinates within the $\n_{\S} - \hat{\tau}$ plane:
\begin{equation}\label{ct2}
\r_{\S} =   (x_{\S} +x_{\S\perp}\frac{{\hat{n}}_{\S\perp}^x}{{\hat{n}}_{\S}^x} ,\tau) = (r_{\S} \cos\theta_{\S}, r_{\S} \sin\theta_{\S})\,,
\end{equation}
because these are the coordinates in which the $\n_{\S} - \hat{\tau}$ planes restricted to each sheet are naturally glued together by extending the range of $\theta_{\S}$ to $2\pi n$, as we will now show.
The shift $x_{\S\perp} \hat{n}_{\S\perp}^x/\hat{n}_{\S}^x$ of $x_{\S}$ is necessary as to ensure that $x=0$, the location of the onset of the branch cut, corresponds to $\r_{\S}=0$ which is
what makes these coordinates convenient.
The $\n_{\S} - \hat{\tau}$ planes are now defined by fixed $x_{\S\perp}$.

If we can establish that the $2\pi n$ periodicity of $\theta$ is equivalent to a $2\pi n$ periodicity of $\theta_{\S}$, then CC's solution would be justified in the above set-up so that the non-interacting Green's function  $G_0(\S,\S; ,r_{\S}, \theta_{\S}, r{'}_{\S}, \theta{'}_{\S})$ can be expressed through CC's result. This can be achieved by establishing a one-to-one correspondence (mapping) between $\theta$ and $\theta_{\S}$. The mapping is intuitively constructed, as shown in Fig. \ref{subfig:bc}, as the vertical projection from the $\n_{\S} - \hat{\tau}$ plane (the blue plane in Fig. \ref{subfig:bc}) onto the $\hat{x} - \hat{\tau}$ plane along
the $\hat y$ direction.
Consider moving the projection of $\r_{\S}$  in the $\hat{x}-\hat{\tau}$ plane around the origin $n$ times (the red circle).
It is clear that $\r_{\S}$ (on the blue ellipse) follows its projection while also moving around the origin $n$ times, always being on the same sheet.
In particular, the branch cut is always traversed simultaneously for $\theta=\theta_{\S}=\pi \mbox{ mod } 2\pi$.
The $\n_{\S} - \hat{\tau}$ planes, the leaves of our foliation, thus have the familiar 1+1$d$ $n$-sheeted geometry,
and $\theta_{\S}$ obeys the same $2n\pi$ periodicity as $\theta$.

Finally, the periodicity condition of the $\hat{y}$ direction is necessary for the total entanglement entropy to be finite;
it also provides the only length scale for the subsystem which is needed for extracting the scaling behavior of entanglement entropy.
However, if we are only concerned with the integral form of the entanglement entropy as in Eq. (\ref{eq:GK}), not requiring it to be finite as a whole, but rather requiring only
the entanglement entropy per unit length to be finite, we may take the $y$ direction to be infinite.
This point of view will be taken here and in the following in order to simplify our calculation.

\subsection{Entanglement entropy of free fermions revisited}
\label{subsec:ee_of_free_fermion_rev}
In order to treat the interacting theory, in this section we re-derive the free fermion result for the half cylinder geometry via the replica trick. Later we shall generalize the method to include interactions. Rewriting the Hamiltonian Eq. (\ref{eq:bH_free}) in terms of the non-chiral fields, and adding the mass term by hand, we have
\begin{widetext}
\beq
\begin{split}
& H[\phi(\S;\x)] = \sum_{\S} \int d^2x   \frac{2\pi v_F^*}{\Omega V} \Big(
  (\pt_{x_{\S}} \varphi(\S;\x) )^2 + (\pt_{x_{\S}} \chi(\S;\x) )^2 \Big) + \frac{m^2}{2} \varphi(\S;\x) ^2.
\end{split}
\eeq
\end{widetext}
For convenience, we use the Lagrangian formalism and work with the $\varphi(\S;\x)$ representation through the rest of this work.

Switching to imaginary time $t\rightarrow i \tau$, and rescaling the coordinates in the following manner:
\beq\label{eq:rescale}
\tau \rightarrow  \sqrt{\frac{V}{16\pi^3 v_F^* \Omega}} \frac{\tau}{m} , \quad
\x \rightarrow  \sqrt{\frac{\Omega V}{4\pi^2 v_F^*}} \frac{\x}{m},
\eeq
we obtain the following Lagrangian density in the $\varphi$ representation :
\beq
\mathcal{L} = -\frac{m^2}{2}[ (\pt_\tau \varphi(\S;\x))^2 + (\pt_{\S} \varphi(\S;\x))^2 +  (\varphi(\S;\x))^2].
\eeq
Then we can work out the Euler-Lagrangian (E-L) equation of motion. Making use of the E-L equation of motion, we find the Green's function
$G_0^{(n)}(\S,\T;\x,\x') = \ev{T \varphi(\S;\x) \varphi(\T;\y)}_0$
satisfies the following differential equation:
\beq\label{eq:diff_free_G}
\begin{split}
& - \left( \partial_\tau^2   + \pt_{x_{\S}}^2  - 1 \right) G_0^{(n)}(\S,\T;\x,\y)  \\ & = C \delta_{\S,\T} \delta(\tau-\tau_y) \delta(x_{\S} - y_{\S})  \delta^{d-1}(\x_{\S \bot} - \y_{\S \bot}),
\end{split}
\eeq
where $C=  2\pi\Omega m^{d-1} \left(\sqrt{\frac{\Omega V}{4\pi^2 v_F^*}}\right)^{d-2} $.
The rescaling makes the Green's function dimensionless, thus easier to handle when it comes to computing $\int d^{d+1} x G^{(n)}(\x,\x)$. The extra factor $C$ generated on the right hand side ($rhs$) will be canceled by the Jacobian of the integral over the Green's function, leaving only a factor of $1/m^2$. All that needs to be computed is then an integral over the dimensionless $G$. Therefore, it is legitimate to ignore this factor from now on. The $\delta$-functions originate from the commutator Eq. (\ref{eq:bcom}), and are {\it coarse-grained}. After we include the patch index, perform the integral over $m^2$, and take the $n$ derivative, Eq. (\ref{eq:entropy}) becomes
\beq\label{eq:entropy_reduced}\begin{split}
& S_A  =   \frac{1}{2} \log (m^2 a_0^2)  \lim_{n \rightarrow 1} \frac{\partial}{\partial n} \sum\limits_{\S} (C_G(\S; n) - n C_G(\S;1)),
\end{split}
\eeq
where $a_0$ is an ultraviolet cutoff, and
\beq C_G(\S; n) = \int d^{d+1} x G_0^{(n)}(\S,\S; \x,\x).\eeq
 The exponential factor in Eq. (\ref{eq:entropy}) becomes one after the $n\rightarrow 1$ limit is applied. Note that $\frac{1}{2} \log (m^2 a^2) \sim - \log L$. Our major task is now computing $C_G(\S;n)$.

Observing that there is no $x_{\S\bot}$ dependence on the left hand side ($lhs$) of Eq. (\ref{eq:diff_free_G}), we can write
$$G_0^{(n)}(\S,\T;\x,\y) = \delta_{\S,\T} \delta^{d-1}(\x_{\S \bot} - \y_{\S\bot}) G_{0,b}^{(n)}(\S;\r_{\x},\r_{\y}),$$ and we obtain a $(1+1)d$ equation
\beq\begin{split}
& -(\partial_\tau^2 + \partial_{x_{\S}}^2 + 1) G_{0,b}^{(n)}(\S;\r_{\S,\x},\r_{\S,\y}) \\ & = \delta(\tau-\tau_y) \delta(x_{\S} - y_{\S})
\end{split}\eeq
in which $\r_{\S,\x(\y)}$ is as defined in Eq. \eqref{ct2}. The same equation appears in CC. We shall also suppress the subscript ${\S}$ unless necessary, as it is normally already specified in the
notation for $G_{0,b}$.

The transverse part of the integral in $C_G(\S;n)$ can be factored out as $$\int d^{d-1}x_{\S\bot} \delta^{d-1}(\x_{\S\bot} - \y_{\S\bot}) \Big\vert_{\y \rightarrow \x}.$$
Recalling our discussion about Eq. (\ref{eq:bcom}), this is a coarse-grained $\delta$-function. At short distances, instead of a divergence, we should use
\begin{equation}\label{ASregularize}
\delta^{d-1}(x_{\S\bot} - y_{\S\bot}) \Big\vert_{\y \rightarrow \x} = (\Lambda/(2\pi))^{d-1}.
\end{equation}
Therefore, the transverse direction integral becomes
$$(\Lambda/(2\pi))^{d-1} \int d^{d-1}\x_{\S\bot} = (\Lambda/(2\pi))^{d-1} \oint_{\partial A} d\S_{\x} \cdot \n_{\S}.$$
Identifying $\Lambda^{d-1} \n_{\S}$ as the surface element $d\S_{\k}$, for a given patch the integration can be rewritten as $(2\pi)^{-d+1} \oint_{\partial_A} \abs{d\S_{\x} \cdot d\S_{\k}}$. This leaves us with only an integral over $(G_{0,b}^{(n)}(\S;\r_x,\r_x) - n G_{0,b}^{(1)}(\S;\r_x,\r_x))$.

The solution for the $(1+1)d$ Green's function on the $n-$sheeted replica manifold is given in CC:
\beq\label{eq:(1+1)d_G}
\begin{split}
G_{0,b}^{(n)}(\S;\r_x,\r_y)
& =  \frac{1}{2\pi n} \sum_{k=0}^{\infty} d_k \mathcal{C}_{k/n}(\theta_x-\theta_y) \\ \times g_{k/n}(r_x,r_y),
\end{split}
\eeq
where $d_0 = 1, d_k = 2$ for $k>0$, $\mathcal{C}_\nu(\theta) = \cos(\nu \theta)$, $g_{\nu}(r,r') = \theta(r - r')  I_{\nu}(r') K_{\nu}(r) + \theta(r' - r) I_{\nu}(r) K_{\nu}(r') $, and $I_\nu(r)$ and $K_\nu(r)$ are the modified Bessel functions of the first and second kind respectively. $r$ and $\theta$ are again the polar coordinates of the  $\n_{\S}-\hat{\tau}$ plane, and we have suppress the index $\S$ of $r$ as only one patch direction is involved.

The integral over $G_{0,b}^{(n)}$ is
\beq
\begin{split}
& \int d^2r_x G_{0,b}^{(n)}(\S;\r_x,\r_x) 
= \int dr_x r_x \sum_k d_k g_{k/n}(r_x,r_x).
\end{split}
\eeq
The integral is divergent since the integrand $r_x g_{k/n}(r_x,r_x)\vert_{r_x\rightarrow \infty} $$= 1/4$, a consequence of the fact that we are calculating the partition function of an infinite system. But this divergence should be canceled in $C_G(\S;n) - n C_G(\S;1)$. To regularize the divergence, we use the Euler-MacLaurin (E-M) summation formula following CC, and sum over $k$ first:
\beq\label{eq:eml_sum}\begin{split}
 \frac{1}{2} \sum_{k=0}^\infty d_k f(k) & = \int_0^\infty f(k) dk - \frac{1}{12} f'(0)  \\
& - \sum_{j=2} ^\infty \frac{B_{2j}}{(2j)!} f^{(2j-1)}(0),
\end{split}\eeq
where $B_{2n}$ are the Bernoulli numbers, $f^{(2j-1)}(0) =
\partial_k^{2j-1} f(k)|_{k=0}$. Note that the first term, the integral over $k$, is always canceled by rescaling $k/n\rightarrow k$ in $g_{k/n}$ .
For the remaining terms, which contain derivatives with respect to $k$,
we may add a constant, in this case $-1/4$, under the derivative,  which allows us to pull the derivative outside the integral.
The integrand now is well-behaved at infinity. To be more precise, according to Eq. (\ref{eq:eml_sum}), we need to compute
\begin{equation}
  \begin{split}
    & \int dr_x r_x \partial_{k}^j g_{k/n}(r_x,r_x) \Big\vert_{k
      \rightarrow 0} \\
    &= \partial_{k}^j \int dr_x ( r_x g_{k/n}(r_x,r_x) - 1/4)\Big\vert_{k
      \rightarrow 0}\\
    &= \partial_{k}^j (-\frac{k}{2n}).
  \end{split}
\end{equation}
So we have
\beq
C_G(\S;n) - n C_G(\S;1) = \frac{1 - n^2}{24 n} .
\eeq
Combining the above results into Eq. (\ref{eq:entropy_reduced}) and converting the sum over $\S$ into an integral around the Fermi surface, we obtain Eq. (\ref{eq:GK}) for this geometry.

\subsection{Differential Equations of the Green's Functions and an Iterative Solution}
In this part, we derive the differential equations of the Green's functions for the quadratic boson theory with inter-patch coupling, and provide an iterative solution. Including the Fermi liquid interaction $V(\S,\T;\x-\y) = U_{\S,\T}$, the Hamiltonian becomes
\beq\begin{split}
& H[\phi(\S;\x)]
= \frac{2\pi v_F^*}{\Omega V} \int d^2x \Big[ \sum_{\S} (\pt_{\S} \phi(\S;\x))^2 \\ & + \sum_{\S,\T} g_{\S,\T} \pt_{\S} \phi(\S;\x) \pt_{\T} \phi(\T;\x) \Big]
\end{split}\eeq
where $g_{\S,\T} = \frac{U_{\S,\T} \Omega}{2\pi v_F^*} $ is order $1/N$. This Hamiltonian can be written in terms of the non-chiral fields as
\begin{widetext}
  \begin{equation}
    H=\frac{2\pi v_F^*}{\Omega V} \int d^2x \Big(\sum_{\S} \big((
    \pt_{x_{\S}} \varphi(\S;\x) )^2 + (\pt_{x_{\S}} \chi(\S;\x) )^2\big)
    + \sum_{\S,\T} g_{\S,\T} (\pt_{x_{\S}}\chi(\S;\x)
    \pt_{x_{\T}}\chi(\T;\x) + \pt_{x_{\S}}\varphi(\S;\x)
    \pt_{x_{\T}}\varphi(\T;\x)) \Big),
  \end{equation}
\end{widetext}
where we have made use of the fact that $g_{\S,\T} = g_{-\S,-\T}$, which is required by time-reversal symmetry.
Here the summation over $\S$ is restricted to a semicircle. This Hamiltonian contains generalized type kinetic terms (inter-patch coupling due to interaction) which are not diagonal. To obtain the corresponding Lagrangian, one needs to invoke the general Legendre transformation\cite{book:goldstein}, and obtains the following Lagrangian densities, respectively,  in terms of $\varphi$ or $\chi$:
\begin{widetext}
  \beq
\begin{split}
  \mathcal{L}_\varphi &= \frac{1}{2} \Big[\sum_{\S}\Big((\pt_t  \varphi(\S;\x))^2 - (\pt_{\S}  \varphi(\S;\x))^2\Big) + \sum_{\S,\T}\Big( h_{2}(\S,\T)\pt_t \varphi(\S;\x) \pt_t \varphi(\T;\x) -  f_{1}(\S,\T)\pt_{\S} \varphi(\S;\x) \pt_{\T} \varphi(\T;\x) \Big) \Big]\\
\mathcal{L}_\chi &= \frac{1}{2} \Big[\sum_{\S}\Big((\pt_t  \chi(\S;\x))^2 - (\pt_{\S}  \chi(\S;\x))^2\Big) + \sum_{\S,\T}\Big( h_{1}(\S,\T)\pt_t \chi(\S;\x) \pt_t \chi(\T;\x) -  f_{2}(\S,\T)\pt_{\S} \chi(\S;\x) \pt_{\T} \chi(\T;\x) \Big) \Big],
\end{split}
\eeq
\end{widetext}
where
\beqarr
\nonumber
&f_1(\S,\T) = g_{\S,\T} + g_{-\S,-\T} - g_{\S,-\T} - g_{-\S,\T}, \\
\nonumber
& f_2(\S,\T) = g_{\S,\T} + g_{- \S,-\T} + g_{-\S,\T} + g_{\S,-\T},
\eeqarr
and $h_{1(2)}(\S,\T)$ is defined through
\begin{equation}
 \{I + [f_{1(2)}(\S,\T)] \}^{-1} = I + [h_{1(2)}(\S,\T)].
\end{equation}
Here, $I$ is the identity matrix, and $[f(h)_{i}(\S,\T)]$ is the matrix formed by $f(h)_{i}(\S,\T)$, $i = 1,2$. Applying this result and making use equations of motion obtained from the Hamiltonian, we obtain the Lagrangians  $\mathcal{L}_\varphi$ or $\mathcal{L}_\chi$. Here we arbitrarily choose to work with $\mathcal{L}\varphi$. Then, by making use of the E-L equation of motion, applying the same rescaling Eq. (\ref{eq:rescale}), and letting $t = i\tau$, we obtain the differential equations that the interacting Green's function $G^{(n)} = \ev{\varphi(\S;\x) \varphi(\T;\x')}$'s satisfies:
\beq\label{eq:G_resc}\begin{split}
&- (\partial_\tau^2 + \pt_{\S}^2  - 1) G^{(n)}(\S,\T;\x,\x') \\ &+ \sum_{l} (h_2(l,\T) \partial_\tau^2 + f_1(l,\T) \pt_{l} \pt_{\T}) G^{(n)}(l,\T;\x,\x') \\
& =  C \delta_{\S,\T} \delta(\tau-\tau') \delta( x_{\S}-x'_{\S})) \delta(x_{\S\bot} - x'_{\S\bot}).
\end{split}
\eeq
Here the Jacobian due to change of variables is the same as in the free fermion case.
The entanglement entropy is still given by Eq. (\ref{eq:entropy_reduced}), but replacing $G_0^{(n)}$ with $G^{(n)}$ in $C_G(\S;n)$. In the following, we omit the replica index $n$ in the Green's function unless different values of $n$ are involved in a single equation.

As is well known, 
differential equations such as the above can be converted to an integral form\cite{Doniach1998} relating
the full Green's function to the noninteracting one. This leads to an iterative (perturbative) definition of the former in terms of the latter.
In the present case, this integral equation reads
\beq\label{eq:iteration_sol}
\begin{split}
& G(\S,\T;\x,\y) \\ & = G_0(\S,\T;\x,\y)   +  \int d^3 z G_0(\S,\S;\x, \z) \\ & \times \Big( \sum_l (h_2(l,\T) \partial_\tau^2 + f_1(l,\T) \pt_{l} \pt_{\T}) G(l,\T;\z,\y) \Big)\\
& = G_0(\S,\T;\x,\y) + \delta G(\S,\T;\x,\y).
\end{split}
\eeq
Given this equation, 
 we can now compute the Green's function and thus the entanglement entropy perturbatively in powers of $U$.

\subsection{Entanglement Entropy from the Iterative Solution}

In Eq. \eqref{eq:iteration_sol}, the $G_0$ term is the same as that of the free fermions, thus yields the same contribution to entanglement entropy. To study how the correction term $\delta G(\S,\T;\x,\y)$ affects the entanglement entropy, we need to study
\begin{equation}
\int d^3x \delta G(\S,\S;\x,\x) = \sum\limits_{M=1}^{\infty} \int d^3x \delta^{(M)}G(\S,\S;\x,\x),
\end{equation}
where $\delta^{(M)}G$ denotes the $M$th order correction.
There are two distinctive types of terms in the perturbative expansion of $\delta G$. In general, at order $M$ , we have in total $3(M+1)$ integrals. Let us examine one of the many terms contributing to the M-th order correction, to be summed over patch indices:
\beq\label{eq:in-patch_O(M)}
\begin{split}
&\int d^3x \delta^{(M)} G(\S,\S;\x,\x) \\
& \sim \int d^3x \prod_{i=0}^{M-1}(d^3z_i) G_0(\S,\S; \x, \z_0) \\
& \times \partial_{\tau_0}^2 G_0(l_0,l_0; \z_0, \z_1)  \dots \times \partial_{\tau_i}^2 G_0(l_i,l_i; \z_i, \z_{i+1}) \\ & \times \dots \times \partial_{\tau_{M-1}}^2 G_0(\S,\S; \z_{M-1}, \x).
\end{split}\eeq
Here we only include the $\tau-$derivatives. In general we would also have spatial ($\n_{l_i}$) derivative terms, as well as terms with mixed derivatives.  But $\hat{\tau}$ and $\n_{\S}$ directions are equivalent. Using rotational symmetry, and the fact that the two different derivatives in each term are with respect to independent variable that are each integrated over, one can see that all  terms are identical except for $S$-dependent pre-factors. The two categories of terms are defined by the set $\{l_i\}$: 1) $l_i = \S$ $\forall\ i$, i.e. with intra-patch coupling only; and 2) $\exists\ l_i \neq \S$ containing inter-patch coupling. We shall label the two categories as
\begin{equation}
  \begin{split}
    & \delta^{(M)}G(\S,\S;\x,\x) = \\
    & \delta^{(M)}_{\text{intra}}G(\S,\S;\x,\x) +
    \delta^{(M)}_{\text{inter}}G(\S,\S;\x,\x)
  \end{split}
\end{equation}

\subsubsection{Intra-patch coupling and comparison with 1D}

Setting $l_i = \S$ for all $i$'s in Eq. \eqref{eq:in-patch_O(M)}, first we consider the transverse direction
$$G_0(\S,\S;\z_i,\z_{i+1}) \sim \delta(z^{(\S)}_{\bot, i} - z^{(\S)}_{\bot,i+1}).$$
We can immediately integrate out the transverse component of all $\z_i$'s and obtain
\begin{equation}
  \begin{split}
    & \int d^3x  \delta^{(M)}_{\text{intra}}G(\S,\S;\x,\x) \sim\\
    & \int dx_{\S\bot} \delta(x_{\S\bot} - z_{\S\bot,0})\delta(z_{\S\bot, M-1} -
    y_{\S\bot})\vert_{\y\rightarrow \x} \\
    &\times \int \prod_{i} dz_{\S\bot,i}
    \prod_{i}  \delta(z_{\S\bot, i} - z_{\S\bot,i+1})\\
    & = \int dx_{\S\bot} \delta(0) = \ L^{d-1} (\Lambda/(2\pi))^{d-1}.
  \end{split}
\end{equation}
In the last line we use again the fact that the transverse $\delta$-function is a coarse-grained one (Eq.\eqref{ASregularize}).

The rest of $\delta^{(M)}_{\text{intra}}G$ is obtained by substituting $G_0(\S,\S;\z_i,\z_{i+1})$ with the $(1+1)d$ Green's function $G_{0,b}(\S;z_{\S,i}, z_{\S,i+1})$.
Although a direct computation is possible, we first give a general argument that for any $M$ the contribution to entanglement entropy from $\delta^M_{\text{intra}} G$ vanishes. We do so by making a comparison with the 1D case where a rigorous solution is available.

For the 1D Luttinger liquid with only forward scattering, the entanglement entropy can be calculated directly via bosonization and the result remains at $1/3\log L$ in the presence of interactions. The calculation is possible because, in our language, there are only two patches, so the transformation which diagonalizes the Hamiltonian is not plagued by the nonlocality issue we encounter in the 2D theory. However, we can also treat the 1D case with our perturbative approach. The resulting series of integrals turns out to be {\em identical} to the one obtained from the intra-patch contributions in the higher dimensional case except for the transverse $\delta-$function. Therefore, we argue that at all orders, the intra-patch coupling terms have vanishing contribution to the entanglement entropy. We shall demonstrate such behavior explicitly up to second order in $U$ later on.

 \subsubsection{Scaling analysis of inter-patch coupling}

For terms with inter-patch coupling, we find that they are of order $\mathcal{O}(1/L)$ comparing to the leading term according a scaling argument.  The crucial observation here is that, as long as $\exists\ l_i \neq \S$, we do not encounter the factor $\delta^{D-1}(0) = L^{D-1} (\Lambda)^{D-1}$, Eq. \eqref{ASregularize} because for $l\neq \S$
\beq
\delta(z_{1\bot} ^{(\S)}- z_{\bot}^{(\S)}) \delta(z_{\bot} ^{(l)}- z_{1\bot}^{(l)}) = \frac{\delta^2(\z_1 - \z)}{\abs{\sin(\theta_l - \theta_{\S})}},
\eeq
where $\theta_{\S}$ ($\theta_l$) is the angle between $\n_{\S}$ ($\n_l$) and the $\hat{x}$-axis in the $x-y$ plane. Therefore, when we integrate out the $(M + 1)$ transverse $\delta$-functions, the factor $\delta^{D-1}(0) = L^{D-1} (\Lambda)^{D-1}$ would be suppressed by even a single $l_i\neq \S$.

To examine the remaining integral, we can ignore the angular part as it cannot affect the scaling behavior. The asymptotic expansion of $K_\nu(r)$ and $I_\nu(r)$ for real $r$  at large value is\cite{wangzhuxi}
\beqarr \nonumber
&K_\nu(r) \simeq \sqrt{\frac{\pi}{2r}} e^{-r} \Big[1 + \sum\limits_{n=1}^{\infty} \frac{(\nu,n)}{(2r)^n}\Big],\\
\nonumber
&I_\nu(r) \simeq \frac{e^r}{\sqrt{2\pi r}} \Big[1 + \sum\limits_{n=1} ^\infty \frac{(-1)^n (\nu,n)}{(2r)^n}\Big],
\eeqarr
where $(\nu,n) = \frac{\Gamma(1/2+\nu+n)}{n! \Gamma(1/2+\nu-n)}$.
By using the above asymptotic expansion of Bessel functions, the leading  term for $\partial_{\tau_0}^2 G_0(l_i,l_; \z_i, \z_{i+1})$ behaves as $\sim \theta(z_i-z_{i+1}) e^{-(z_{i}-z_{i+1})}/(z_{i}-z_{i+1}) +  \theta(z_{i+1}-z_{i})  e^{-(z_{i+1}-z_{i})}/(z_{i+1}-z_{i})$. All of these terms peak around $z_{i+1} = z_{i}$ and are otherwise exponentially suppressed. We may therefore again estimate this integral by letting $\x = \z_0=\z_1=\dots=\z_{M-1}$ and removing $(M+1)$ of the  integrals. The remaining integrals yield, at the leading order,
$\int d^{M+1}z \,1/z^{M} \sim  \int dz z^M z^{-M}$. However, at the leading order, there is no $\nu$ dependence. According to the formalism in Sec. \ref{subsec:ee_of_free_fermion_rev}, such terms have no contribution to the entanglement entropy. Therefore, the term that contributes to the entanglement entropy is the next order which behaves as  $\int dz \frac{1}{z} $ and is of order $\mathcal{O}(\log L)$, leading only to a correction $\sim \mathcal{O}(\log L)\times \log L$ to the entanglement entropy.

Next, we shall demonstrate in detail our above analysis, for both inter-patch and intra-patch coupling terms by explicit calculation up to the second order.

\subsubsection{First Order Correction}
The first order term correction to $\int dx G(\S,\S;\x,\x)$ is
\beq\label{eq:CG1}
\begin{split}
& \delta^{(1)} C_G(\S; n) =  \int d^3x d^3z  G_0(\S,\S;\x,\z)  \\ & \times (h_2(\S,\S) \partial_{\tau_z}^2 + f_1(\S,\S)\pt_{z_{\S}}^2) G_0(\S,\S;\z,\x).
\end{split}
\eeq
As we have pointed, it is sufficient to calculate either piece of the two terms due to the equivalence of the imaginary time direction and the real space direction. The other piece should be just the same except for the coefficient. Here we choose to compute the first term.

The transverse degrees of freedom provide an overall factor counting the total degrees of freedom as discussed in the general case. Then we can also integrate out the angular degrees of freedom in the $\hat{x}_{\S}-\hat{\tau}$ plane, both $\theta_x$ and $\theta_z$ as defined in Eq. (\ref{eq:(1+1)d_G}), after which one obtains
\beq\label{eq:dG1}
\begin{split}
&\delta^{(1)}C_G(\S; n) \sim \sum_k  \frac{d_k}{2} \delta^{(1)} G_{k/n} \oint_{\partial A} \abs{d\S_x\cdot d\S_{k}},
\end{split}
\eeq
where
\beq\begin{split}
& \delta^{(1)} G_{k/n} = \int dr_x dr_z r_x r_z g_{k/n} ( r_x, r_z) \\ & \times(\partial_{r_z}^2 - \frac{k^2}{r_z^2 n^2}) g_{k/n} ( r_x, r_z) .
\end{split}\eeq
The two $k$ summation is reduced to one due to orthogonality of the angular function $\mC_{k/n}(\theta)$. By employing the E-M formula and properties of the Bessel functions,
we show in Appendix  \ref{apdx:A} that sum over $k$-values in Eq. \eqref{eq:dG1} can be converted into an integral, which cancels in Eq. \eqref{eq:entropy_reduced}  for the same scaling reasons
discussed above, following Eq. \eqref{eq:eml_sum}.
Therefore, we find that the contribution of Eq. \eqref{eq:CG1} to the entanglement entropy vanishes. 

\subsubsection{Second Order Correction}
The second order correction is
\beq\begin{split}
& \delta^{(2)}C_G(\S;n) =  \int d^3x d^3z d^3z_1 G_0(\S,\S;\x,\z) \\ & \times   \sum_l  \Big( h_2(l,\S) \partial_{\tau_{z}}^2 + f_1(l,\S) \pt_{z_l}\pt_{z_{\S}}\Big) G_0(l,l;\z,\z_1) \\
& \times (h_2(\S,\S) \partial_{\tau_{z_1}}^2 + f_1(\S,\S) \pt_{z_{1\S}}^2) G_0(\S,\S;\z_1,\x).
\end{split}
\eeq

$\bullet \quad$ for $l = \S$:
\beq\begin{split}
& \int d^3x d^3z d^3z_1 G_0(\S,\S;\x,\z) \\ & \times \big(h_2(\S,\S) \partial_{\tau_{z}}^2 + f_1(\S,\S) \pt_{z_{\S}}^2\big)G_0(\S, \S;\z,\z_1) \\
& \times  (h_2(\S,\S) \partial_{\tau_{z_1}}^2 + f_1(\S,\S) \pt_{z_{1\S}}^2) G_0(\S,\S;\z_1,\x).
\end{split}
\eeq
 According to our general discussion, we only need to consider the following piece:
\beq\begin{split}\label{eq:ksum2}
& \int d^3x d^3z d^3z_1 G_0(\S,\S;\x,\z)  \partial_{\tau_{z}}^2 G_0(\S, \S;\z,\z_1) \\
&  \times \partial_{\tau_{z_1}}^2 G_0(\S,\S;\z_1,\x)  \\
& = (2\pi)^{-1} \oint_{\partial A} \abs{d\S_x\cdot d\S_{k}} \sum_k \frac{d_k}{4} \delta^{(2)}G_{k/n},
\end{split}
\eeq
where
\beq\label{eq:dG2}
\begin{split}
&\delta^{(2)}G_{k/n} = \int dr_x dr_z dr_1 r_x r_z r_1 g_{k/n} (r_x, r_z) \\
& \times ( \partial_{r_z}^2 - k^2/(r_z n)^2)  g_{k/n} ( r_z, r_1)\\
& \times ( \partial_{r_1}^2 - k^2/(r_1 n)^2)  g_{k/n} ( r_1, r_x).
\end{split}\eeq
In the above, we have proceeded as in the first order calculation, integrating out the angular part first to obtain the expression for $\delta^{(2)}G_{k/n} $.

After a lengthy but similar calculation as for the first order (see Appendix B), we find, using the E-M formula:
\beq\begin{split}\label{eq:EM2alex}
&\frac{1}{2} \sum d_k \delta^{(2)}G_{k/n}= \int dr_x dr_z dr_1 r_x r_z r_1\\
& \times \int_0^\infty dk \,p_{k/n}(r_x,r_z,r_1) \\
&+ \left[\frac{1}{12}  \partial _k +\sum_{j=2}^\infty \frac{B_{2j}}{(2j)!}\partial_k^{(2j-1)} \right] \frac{n}{16k},
\end{split}
\eeq
where $p_{k/n}(r_x,r_z,r_1)$ is the product of $g_{k/n}$ dependent terms in Eq. \eqref{eq:dG2}.
The usual scaling argument for the integral shows that the entire expression is proportional to $n$, and
thus cancels the second ($n=1$) term in  Eq. \eqref{eq:entropy_reduced}:
\beq\nonumber
\begin{split}
& S (\S) \sim  - \frac{\partial}{\partial n}\int d^2x (G_n - n G_1) )\bigg\vert_{n=1}.
\end{split}
\eeq
Therefore, at the second order level for the $l = \S$ piece we still have no correction to the scaling law of entanglement entropy.

$\bullet \quad$ for $l \neq \S$:

The integrand we need to consider is
\beq\label{eq:2nd_order}\begin{split}
& G_0(\S,\S;\x,\z) \\ &\times  \Big( h_2(l,\S) \partial_{\tau_{z}}^2 + f_1(l,\S) \pt_{z_l}\pt_{z_{\S}}\Big) G_0(l,l;\z,\z_1)  \\
& \times (h_2(\S,\S) \partial_{\tau_{z_1}}^2 + f_1(\S,\S) \pt_{z_{1\S}}^2) G_0(\S,\S;\z_1,\x).\\
\end{split}
\eeq

The first thing to notice in Eq. (\ref{eq:2nd_order}) is that we have derivatives along directions different from the patch normal direction $\n_{\S}$ acting on the non-interacting Green's function. We expand this term as
\begin{equation}
  \begin{split}
    & \pt_{z_l}\pt_{z_{\S}} G_0(l,l;\z,\z_1) = \delta(z_{\perp}^{(l)} -
    z_{1\perp}^{(l)})\pt_{z_l}\pt_{z_{\S}}G_{0,b}(l;\z,\z_1)\\
    & + \pt_{z_{\S}} \delta(z_{\perp}^{(l)} - z_{1\perp}^{(l)}) \pt_{z_l} G_{0,b}(l;\z,\z_1).
  \end{split}
\end{equation}
For the first term, we can decompose the derivative $\pt_{z_{\S}}$ into terms that act along $\n_{l}$ and along its transverse direction, respectively. The non-interacting Green's function only depends on the transverse coordinates via $ G_0(l,l;\x,\y) \sim \delta(x_{\bot}^{(l)}-y_{\bot}^{(l)})$, which indicates that those derivative terms vanish. Thus it is $\sim \delta(z_{\perp}^{(l)} - z_{1\perp}^{(l)})\pt_{z_l}^2G_{0,b}(l;\z,\z_1)$. For the second term, we integrate by parts with respect to $z_{\S}$, which leads to (including now the first $G_0$ factor, which depends on $\z$)
\begin{equation}\nonumber
 - \delta(x_{\bot}^{(\S)} - z_{1\bot} ^{(\S)})  \delta(z_{\bot} ^{(l)}- z_{1\bot}^{(l)}) \pt_{z_{\S}} G_{0,b}(\S;\x,\z) \pt_{z_l} G_{0,b}(l;\z,\z_1).
\end{equation}
Therefore, the overall integrand is proportional to $ \delta(x_{\bot} ^{(\S)}- z_{\bot}^{(\S)}) \delta(z_{1\bot} ^{(\S)}-x_{\bot}^{(\S)})  \delta(z_{\bot} ^{(l)}- z_{1\bot}^{(l)})$. Note that the $x_{\bot}^{(\S)}$ dependence only appears in these $\delta$-functions, we can integrate it out, leaving only $ \delta( z_{\bot}^{(\S)} - z_{1\bot} ^{(\S)}) \delta(z_{\bot} ^{(l)}- z_{1\bot}^{(l)}) \sim \delta(\z_1 - \z)$.

Secondly, it is sufficient to focus on the following terms in the integrand
\begin{equation}\nonumber
  \begin{split}
    & \Big(G_{0,b}(\S;\x,\z) \partial_{\tau_{z}}^2 G_{0,b}(l;\z,\z_1) +
    \pt_{z_{\S}}G_{0,b}(\S;\x,\z) \\
& \times \partial_{z_l} G_{0,b}(l;\z,\z_1)\Big)   \partial_{\tau_{z_1}}^2 G_{0,b}(\S;\z_1,\x),
\end{split}
 \end{equation}
to ease the presentation. For other combinations, the rest of this section is equally applicable with minor modifications that only leads to different coefficients and do not affects the scaling analysis.
We first perform the intra-patch integration
\beq
 \int d^3x G_{0,b}(\S;\x,\z)  G_{0,b}(\S;\z_1,\x)= H(\S;\z, \z_1),
\eeq
where
\beq\nonumber
    \begin{split}
      & H(\S;\z, \z_1) = \sum_k \frac{d_k}{2\pi n} \mC_{k/n}(\theta_z,\theta_{z_1}) (\theta(r_z-r_1) \\
 &\times (r_z K_{z+} I_1 - r_1 I_{1-} K_z) + \theta(r_1 -r_z) \\ & \times (r_1K_{1+} I_z - r_z I_{z-} K_1) ).
    \end{split}
\eeq

So for a given $\S$ the contribution to entanglement entropy due to coupling with patch $l$ can be written as
\beq\label{eq:2nd_order_scaling}
\begin{split}
& \Big| \frac{1}{\sin(\theta_l - \theta_{\S})}\Big| \int d^3z d^3z_1  \delta^2(\z_1 - \z) \Big(\partial_{\tau_{z_1}}^2  H(\S;\z, \z_1)  \\
&\times \partial_\tau^2 G_{0,b} (l;\z_1,\z)  + \pt_{z_{\S}} \partial_{\tau_{z_1}}^2 H(\S;\z, \z_1) \pt_{z_l} G_{0,b} (l;\z_1,\z)\Big)\\
& =  \Big| \frac{1}{\sin(\theta_l - \theta_{\S})}\Big| \int d^3z d\tau_1  \Big(\partial_{\tau_{z_1}}^2 H(\S;\z, \z_1)  \partial_\tau^2 G_{0,b} (l;\z_1,\z)  \\
& + \pt_{z_{\S}} \partial_{\tau_{z_1}}^2 H(\S;\z, \z_1) \pt_{z_l} G_{0,b} (l;\z_1,\z) \Big) \Big|_{\substack{z_{1,x} = z_{,x}\\ z_{1,y} = z_{,y}}},
\end{split}
\eeq
where $z_{,x}, z_{,y}$ indicate the two spatial components of $\z$.

%
As we argued in previous section,  for extracting the order of magnitude of the result it is sufficient to set $\tau_1 = \tau$ in the final line of Eq. (\ref{eq:2nd_order_scaling}) and remove the integral over $\tau_1$. We also note that the derivatives do not alter the leading power of $r$, owing to the presence of the exponential function. Therefore, it is sufficient to examine
\beq
\int d^3z \left( H(\S;\z, \z_1)  G_{0,b} (l;\z_1,\z)\right)\Big|_{\z_1 = \z}.
\eeq
At the lowest order in $1/r$, we have
\beq
G_{0,b}(\S;r,r) \sim I_\nu(r) K_\nu(r) \sim 1/r,
\eeq
\beq\begin{split}
&H(\S;r,r) \sim r I_\nu(r) K_{\nu+}(r) - r K_\nu(r) I_{\nu-}(r) \\
= &(1+\frac{(\nu + 1,1)}{2r}+\dots) (1 - \frac{(\nu, 1)}{2r} + \dots)\\
& -  (1-\frac{(\nu - 1,1)}{2r}+\dots) (1 + \frac{(\nu, 1)}{2r} + \dots)\\
= &\frac{1}{r} + \mathcal{O}(\frac{1}{r^2}).
\end{split}
\eeq
Since the $\tau$ derivative does not alter the leading powers, we extract the leading term to be
\beq
\left( H(\S;\z, \z_1)  \partial_\tau^2 G_{0,b} (\S;\z_1,\z) \right)\Big|_{\z_1 = z} \sim \frac{1}{z^2}.
\eeq
For a triple integral over $1/z^2$, one would get a linear divergence, i.e. the result would be $\sim L$. This is indeed the case as we have already seen in previous calculation. However, at the lowest order, everything is independent on $\nu = k/n$. Actually what finally appears in the the entanglement entropy are the $k-$derivatives of these terms appearing in the E-L summation formula. This means the leading term has vanishing contribution to the entanglement entropy. The first term contributing to entanglement entropy is then $\sim \int d^3z \frac{1}{z^3}$ the upper limit of which is  order $\mathcal{O}(\log L)$ and only leads a correction up to $\sim \mathcal{O}(\log L)\times \log L$ to the free fermion entanglement entropy.

\section{Summary and Concluding Remarks}

In this paper, we developed an intuitive understanding of the logarithmic correction to the area law for the entanglement entropy of free fermions in one and higher dimensions on equal footing -- the criticality associated with the Fermi surface (or points). Then we used the tool of high dimensional bosonization to compute the entanglement entropy, and generalized this procedure to include Fermi liquid interactions. In the presence of such interactions we calculated the entanglement entropy for a special geometry perturbatively in powers of the interaction strength up to the second order, and find no correction to the leading scaling behavior. We also point out that the situation is the same at higher orders. Our results thus strongly suggest that the leading scaling behavior of the block entanglement entropy of a Fermi liquid is the {\em same} as that of a free Fermi gas with the same Fermi surface, not only for the special block geometry studied in this paper, but for arbitrary geometries. Explicit demonstration of the latter is an obvious direction for future work.

In the special geometry in which we performed explicit calculations using the replica trick, a mass-like term is introduced to regularize the theory at long distance, as is done in closely related contexts\cite{Korepin2004,Hertzberg2011}. For a Fermi liquid (which is quantum-critical) the corresponding length scale $\xi\sim v_F/m$ must be identified with the block size $L$, and is thus {\em not} an independent length scale. On the other hand, such a mass-like term can also describe a superconducting gap due to pairing. In particular, for a weak-coupling superconductor, $\xi$, the superconducting coherence length, is much longer than all microscopic length scales, but finite nevertheless. In this case it is {\em independent} of $L$, and the interplay between the two is interesting. For $L < \xi$, the Fermi liquid result (\ref{eq:toy_model}) still holds. But for $L > \xi$, the logarithmic factor in the entanglement entropy {\em saturates} at $\log\xi$, and we expect:
\begin{equation}
  \label{eq:finite_xi}
S(\rho_A) =  \frac{1}{12 (2\pi)^{2-1}} \log \xi \times
\oint_{\partial A} \oint_{\partial \Gamma} \abs{dS_x \cdot dS_k},
\end{equation}
which agrees with the conjecture made in Ref.\cite{Swingle2010}.

More generally,  Fermi liquids are (perhaps the best understood) examples of quantum critical phases (or points) in high dimensions. Unlike in 1D where conformal symmetry powerfully constrains the behavior of entanglement entropy, our understanding of entanglement properties of such high-dimensional quantum critical phases or points (many of them have Fermi surfaces but are {\em not} Fermi liquids) is very limited. Our work can be viewed as a step in that general direction.
Furthermore, the formalism developed in this work has potential applicability to systems with composite or emergent fermions with Fermi surfaces as well, or more generally, {\em non-Fermi liquid} phases with Fermi surfaces. The system studied in Ref.\cite{YZhang2011}, where there is an emergent spinon Fermi surface, is a potential example.

\begin{acknowledgements}
 This work is supported support by National Science Foundation under NSF Grant No. DMR-0704133 and DMR-1004545 (W.X.D. and K.Y.). AS is supported  by the National Science Foundation under NSF Grant No. DMR-0907793.
\end{acknowledgements}

\appendix
\section{Calculation of $\delta^{(1)} G_{k/n}$}\label{apdx:A}
Throughout the Appendix, we shall denote the modified Bessel functions $K_\nu(r_i)$, $I_\nu(r_i)$ as $K_i$, $I_i$ for simplicity with $\nu=k/n$. We also have $K(I)_{\nu \pm 1}(r_i)$ which shall be shortened as $K(I)_{i,\pm}$.
\beq\begin{split}
& \delta^{(1)} G_{k/n} = \int dr_x dr_z r_x r_z g_{k/n} ( r_x, r_z) \\
& \times (\partial_{r_z}^2 - \frac{k^2}{r_z^2 n^2}) g_{k/n} ( r_z, r_x).
\end{split}\eeq
Expanding $ (\partial_{r_z}^2 - \frac{k^2}{r_z^2 n^2}) g_{k/n} ( r_z, r_x) $, and noting the identities $I'K - K'I = 1/x $, $X''-(\nu^2/x^2) X = X - (1/x) X'$ where $X= K, I$,
 we get
\beq\begin{split}
& (\partial_{r_z}^2 - \frac{k^2}{r_z^2 n^2}) g_{k/n} ( r_z, r_x) \\
=& - \frac{\delta(r_x-r_z)}{r_x} + \theta(r_x - r_z) (1 - \frac{1}{r_z}\partial_{r_z})I_zK_x \\ &+ \theta(r_z - r_x) (1 - \frac{1}{r_z}\partial_{r_z})I_xK_z.
\end{split}
\eeq
Then integrating over $r_x$ first, and making use of the following formula
\beq
\int dx x X_\nu^2(x) = \frac{1}{2} x^2 (X_\nu^2(x) - X_{\nu - 1}(x)
X_{\nu+1} (x)),
\eeq
where $X_\nu(x)$ can be the first or second kind of modified Bessel function, $I_\nu$ or $K_\nu$,  and the identities $I_-K+IK_- = I_+K + I K_+ =1/x$ in addition to those given above, $\delta^{(1)} G_{k/n}$ is reduced to
\beq\label{eq:simplified_dG}
\begin{split}
 & \delta^{(1)} G_{k/n} =  \int dr r\Big(-  I K + r^2 (I^2-I_+I_-) K^2\Big).
\end{split}
\eeq
We apply the same strategy as in Sec. \ref{subsec:ee_of_free_fermion_rev}, making use of the E-M formula to do the sum over the $k$-index in Eq. \eqref{eq:dG1}.
This converts the sum into a divergent integral over $k$ which cancels in Eq. \eqref{eq:entropy_reduced} as before, and a sum over terms of the form
$\pt_k^j \delta^{(1)} G_{k/n}\big\vert_{k\rightarrow 0}$ that turn out to vanish, as we will now show.
Again, we can include proper constants under the derivative into the integrand. These derivatives then act on well defined integrals.
 The first term has been discussed in Sec. \ref{subsec:ee_of_free_fermion_rev}:
\begin{equation}
  \begin{split}
   &  \int dr \pt_k^j (- r I_{k/n} K_{k/n}) \\
& = \pt_k^j \int dr (- r I K + \frac{1}{4}) = \pt_k^j \Big(\frac{k}{2n}\Big).
  \end{split}
\end{equation}
The second term can be shown to be
\begin{equation}
  \begin{split}
   &  \int dr \pt_k^j (r^3 (I^2-I_+I_-) K^2) \\
& = \pt_k^j \int dr (r^3 (I^2-I_+I_-) K^2 - \frac{1}{4}) \\
& = \pt_k^j \Big(-\frac{1}{16}-\frac{k}{2n}\Big).
  \end{split}
\end{equation}
Summing the two terms together, we find
\beq
\Big(\pt_k^j \delta^{(1)} G_{k/n}\Big)\Big\vert_{k\rightarrow 0} = \pt_k^j (-\frac{1}{16}) =0,
\eeq
for all $j>0$. 

\section{Calculation of $\delta^{(2)}G_{k/n}$}
Let us first compute the integral:
\beq
\begin{split}
& \int  dr_2 r_2 g_{\nu}(r, r_2) \Big(\partial_{r_2}^2g_{\nu}(r_2,r_1) - \left(\frac{\nu}{r_2}\right)^2 \\
& \times g_{k/n}(r_2, r_1) \Big) \\
& = \theta(r - r_1) h(r,r_1) +  \theta(r_1 - r) h(r_1,r)
\end{split}
\eeq
with
\beq\nonumber \begin{split}
&h(r,r_1) = - I_1K + \frac{1}{2} (f_1(r,r_1) +KI_1 \ln \frac{r}{r_1}),\\
&f_1(r,r_1)   = K K_1 (r_1^2 (I_1^2-I_{1,+} I_{1,-}) -I_1^2) - I I_1 \\ & \times (r^2 (K^2 - K_+ K_-) - K^2) + K I_1 (F(r) - F(r_1) \\ & - I K + I_1 K_1),\\
&F(r) = 2 \int dr r I K =  r^2 I K + r^2 I_+ K_-.\end{split}\eeq
The $IK_1$ term in $h(r,r_1) $ results from $\delta$-functions (derivatives of $\delta$-functions) coming from the derivative applied on the step function ($\theta(r)$). The remaining part comes from terms involving a product of two $\theta$-functions. Here one needs to distinguish between $r > r_1$ and $r < r_1$, which gives rise to the terms in $\theta(r - r_1)$ and $\theta(r_1 - r)$, respectively.
 The remaining integral
\beq
\begin{split}\label{eq:remaining}
&\int dr dr_1  r r_1 ( \theta(r - r_1) h(r,r_1) +  \theta(r_1 - r) h(r_1,r) ) \\
&\times \Big(-\frac{\delta(r_1-r)}{r} + \theta(r_1-r) (1-\frac{1}{r_1}\partial_{r_1})K_1 I \\ &+ \theta(r-r_1) (1-\frac{1}{r_1} \partial_{r_1})K I_1\Big) \\
\end{split}
\eeq
can be carried out by applying the identities of Bessel functions $I$ and $K$ used in Appx. A.
In applying the E-M formula to the sum over $k$ in \eqref{eq:ksum2}, we again arrive at a divergent
k-integral that can be rescaled and subsequently canceled (see \eqref{eq:EM2alex} and below),
and a sum over derivative terms that are well-behaved. In the latter terms, we always add proper constants
under the derivatives to regularize the integrand at infinity, as before.
We divide \eqref{eq:remaining}  into two terms.
The first is the one containing the $\delta$-function. After integrating out $r_1$, this term becomes
\beq\begin{split}
 &\pt_k^j (- \int dr r h(r,r) ) = \pt_k^j (- \int dr (r h(r,r) + \frac{1}{4}) ) \\
& = \pt_k^j ( - \int dr (r (-IK + \frac{r^2}{2} (- K^2 I_+I_- \\
& + I^2K_+K_-)) + \frac{1}{4}) )= \pt_k^j (0).
\end{split}
\eeq
The second term is expanded to
\begin{equation}
\begin{split}
& \int dr dr_1 r r_1 \Big( \theta(r-r_1) h(r,r_1) K (I_1 - I_1'/r_1) \\ &+ \theta(r_1-r) h(r_1,r) I (K_1 - K_1'/r_1) \Big).
\end{split}
\end{equation}
Due to the complexity of $h(r,r_1)$, we examine each of the three terms of $h(r,r_1)$ separately. The first term is simple. Applying those identities of $K$'s and $I$'s and including the proper constant, we get
\beq\label{eq:h1}\begin{split}
& \pt_k^j \Big(\int dr dr_1 \big(r r_1 ( \theta(r-r_1) (-I_1 K)K (I_1 - I_1'/r_1) \\ &+ \theta(r_1-r) (-I K_1) I (K_1 - K_1'/r_1) ) - \frac{1}{4}\big)\Big)\\
&= \pt_k^j \Big( \frac{k}{2n} \Big).
\end{split}
\eeq
The second term of $h(r,r_1)$ contributes
\beq\label{eq:h2}
\begin{split}
&\pt_k^j \Big( \frac{1}{2}\int dr dr_1 r r_1 \theta(r - r_1) f1(r,r_1) K
  (I_1 - \frac{1}{r_1} I_1') \\ & + \frac{1}{2} \int dr dr_1 r r_1 \theta(r_1 - r)
  f1(r_1,r) (K_1 -  \frac{1}{r_1} K_1') I \Big).\\
\end{split}
\eeq
By interchanging the dummy variables $r$ and $r_1$, employing the properties of the modified Bessel functions with care, and including the regularization constant, we arrive at
\beq\begin{split}
& \pt_k^j  \Big ( \frac{1}{2}\int dr dr_1 \big( r r_1 \theta(r - r_1) f_1(r,r_1) (2 K I_1 - K I_1'/r_1 \\ &-  K' I_1/r) -  \frac{1}{8}\big) \Big)= \pt_k^j  (-\frac{k}{2n} ).
\end{split}
\eeq
The last part of $h(r,r_1)$ contributes as
\beq\begin{split}
&\int dr dr_1 \frac{r r_1}{2} \theta(r - r_1) K I_1 \ln \frac{r}{r_1} (
 K' I_1 /r  \\ & - K I_1' /r_1) = - \frac{1}{16\nu}.
\end{split}
\eeq

Summing all the above terms together we get
\beq
\pt_k^j (\delta^{(2)}G_{k/n})  =  \pt_k^j (- \frac{n}{16 k} ).
\eeq 
\bibliographystyle{apsrev}

\end{document}